\documentclass[12pt,preprint]{aastex}
\usepackage{color}

\newcommand{\degree}{{^\circ}}

\newcommand{\cm}{{\rm\,cm}}

\newcommand{\au}{{\rm\,AU}}
\newcommand{\AU}{\au}

\newcommand{\gm}{{\rm\,g}}
\newcommand{\gram}{{\gm}}
\newcommand{\kms}{{\rm\,km\,s^{-1}}}

\newcommand{\msun}{{\rm\,M_\odot}}

\newcommand{\second}{{\rm\,s}}

\newcommand{\muG}{{\,\mu\rm G}}
\newcommand{\etaunit}{{\cm^2\second^{-1}}}

\newcommand{\ct}{\citealt}

\begin{document}

\title{Does Magnetic Field-Rotation Misalignment Solve the Magnetic
  Braking Catastrophe in Protostellar Disk Formation?}

\author{Zhi-Yun Li\altaffilmark{1,3}, Ruben Krasnopolsky\altaffilmark{2,3}, Hsien Shang\altaffilmark{2,3}}
\altaffiltext{1}{University of Virginia, Astronomy Department, Charlottesville, USA}
\altaffiltext{2}{Academia Sinica, Institute of Astronomy and Astrophysics, Taipei, Taiwan}
\altaffiltext{3}{Academia Sinica, Theoretical Institute for Advanced Research in Astrophysics, Taipei, Taiwan}

%\shortauthors{{\sc Li et al.{} }}
\shorttitle{\sc Magnetic Field-Rotation Misalignment and Disk Formation}

\begin{abstract}

Stars form in dense cores of molecular clouds that are observed to be
significantly
magnetized. In the simplest case of a laminar (non-turbulent) core
with the magnetic
field aligned with the rotation axis, both analytic considerations
and numerical simulations have shown that the formation of a large,
$10^2\au$-scale, rotationally supported protostellar disk is suppressed
by magnetic braking in the ideal MHD limit for a realistic level
of core magnetization. This theoretical difficulty in forming
protostellar disks is termed ``magnetic braking catastrophe.'' A
possible resolution to this problem, proposed by
\citeauthor{HennebelleCiardi2009} and \citeauthor{Joos+2012},
is that misalignment between the magnetic field and
rotation axis may weaken the magnetic braking enough to enable disk
formation. We evaluate this possibility quantitatively through
numerical simulations. We confirm the basic result of \citeauthor{Joos+2012}
that the misalignment
is indeed conducive to disk formation. In relatively weakly magnetized
cores with dimensionless mass-to-flux ratio $\gtrsim 5$, it
enabled the formation of rotationally supported disks that would
otherwise be suppressed if the magnetic field and rotation axis are
aligned. For more strongly magnetized cores, disk
formation remains suppressed, however, even for the maximum tilt angle
of $90\degree$. If dense cores are as strongly magnetized as indicated
by OH Zeeman observations (with a mean dimensionless mass-to-flux
ratio $\sim 2$), it would be difficult for the misalignment alone
to enable disk formation in the majority of them. We conclude that,
while beneficial to disk formation, especially for the relatively
weak field case, the misalignment does not completely solve the
problem of catastrophic magnetic braking in general.
\end{abstract}

\section{Introduction}

Star and planet formation are connected through disks. Disk formation,
long thought to be a trivial consequence of angular momentum
conservation during core collapse and star formation (e.g.,
\ct{Bodenheimer1995}), turned out to be much more complicated than
originally envisioned. The complication comes from magnetic fields,
which are observed in dense, star-forming, cores of molecular
clouds (see \ct{Crutcher2012} for a recent review). The field can
strongly affect the angular momentum evolution of core collapse
and disk formation through magnetic braking.

There have been a number of studies aiming at quantifying the
effects of magnetic field on disk formation. In the ideal MHD
limit, both analytic considerations and numerical simulations
have shown that the formation of a rotationally supported disk
(RSD hereafter) is suppressed by a realistic magnetic field
(corresponding to a
dimensionless mass-to-flux ratio of $\lambda \sim$ a few;
\ct{TrolandCrutcher2008})
during the protostellar mass accretion phase in
the simplest case of a non-turbulent core with the magnetic field
aligned with the rotation axis (\ct{Allen+2003}; \ct{Galli+2006};
\ct{PriceBate2007}; \ct{MellonLi2008}; \ct{HennebelleFromang2008};
\ct{DappBasu2010}; \ct{Seifried+2011}; \ct{Santos-Lima+2012}).
The suppression of RSDs by excessive magnetic braking is termed
``magnetic braking catastrophe'' in star formation.

Rotationally supported disks are routinely observed, however, around
evolved Class II young stellar objects (see \ct{WilliamsCieza2011}
for a review), and increasingly around Class I (e.g.,
\ct{Jorgensen+2009}; \ct{Lee2011}; \ct{Takakuwa+2012})
and even one Class 0 source (\ct{Tobin+2012}).
When and how such disks form in view of the magnetic braking
catastrophe is unclear. The current attempts to overcome the
catastrophic braking fall into
three categories: (1) non-ideal MHD effects, including ambipolar
diffusion, Ohmic dissipation and Hall effect, (2) misalignment
between magnetic and rotation axes, and (3) turbulence. Ambipolar
diffusion does not appear to weaken the braking enough to enable
large-scale RSD formation under realistic conditions (\ct{KrasnopolskyKonigl2002};
\ct{MellonLi2009}; \ct{DuffinPudritz2009};
\ct{Li+2011}). Ohmic dissipation can produce small, AU-scale,
RSD in the early protostellar accretion phase (\ct{Machida+2010};
\ct{DappBasu2010}; \ct{Dapp+2012}; \ct{Tomida+2013}). Larger, $10^2\AU$-scale
RSDs can be produced if the resistivity or the Hall coefficient
of the dense core is
much larger than the classical (microscopic) value (\ct{Krasnopolsky+2010,Krasnopolsky+2011};
see also \ct{BraidingWardle2012a,BraidingWardle2012b}). \citet{Joos+2012}
explored the effects of tilting the magnetic field away from
the rotation axis on disk formation (see also \ct{Machida+2006};
\ct{PriceBate2007};
\ct{HennebelleCiardi2009}). They concluded that
Keplerian disks can form for a mass-to-flux ratio $\lambda$ as low
as $3$, as long as the tilt angle is close to $90\degree$ (see their
Fig.\ 14). The effects of turbulence were explored by \citet{Santos-Lima+2012,Santos-Lima+2013},
who concluded that a strong enough turbulence can
induce enough magnetic diffusion to enable the formation of a
$10^2\AU$-scale RSD.  \citet{Seifried+2012} and \citet{Myers+2012}
considered supersonically turbulent massive cores. They found
rotationally dominated
disks around low-mass stars,
although in both cases the turbulence-induced rotation is
misaligned with the initial magnetic field by a large angle,
which may have contributed to the disk formation (see also \ct{Joos+2013}).

The goal of this paper is to revisit the role of magnetic
field-rotation misalignment in disk formation. The misalignment
is expected if the angular momenta of dense cores are generated
through turbulent motions (e.g., \ct{BurkertBodenheimer2000};
\ct{Myers+2012}). It is also inferred from the misalignment between
the field direction traced by polarized dust emission and the
outflow axis, which is taken as a proxy for the direction
of rotation (\ct{Hull+2012}). Indeed, in the
CARMA sample of \citeauthor{Hull+2012}, the distribution of the angle $\theta_0$
between the magnetic field and jet/rotation axis is consistent with
being random. If true, it would indicate that in half of the sources
the two axes are misaligned by a large angle of $\theta_0 > 60\degree$
(see however \ct{Chapman+2013} and discussion in \S\ref{disk}).
Such a large misalignment would be enough to allow disk formation
in dense cores magnetized to a realistic level (with $\lambda$ of
a few; \ct{TrolandCrutcher2008}) according to \citet{Joos+2012}.
If the alignment angle $\theta_0$
is indeed random and \citeauthor{Joos+2012}'s conclusions are generally true,
the magnetic braking catastrophe would be largely solved. Given their
far-reaching implications, it is prudent to check \citeauthor{Joos+2012}'s
conclusions, using a different numerical code. It is the task of
this paper.

We carry out numerical experiments of disk formation in dense cores
with misaligned magnetic and rotation axes using non-ideal MHD code
Zeus-TW that includes self-gravity. We find that a large misalignment 
angle does indeed enable the formation of RSDs in weakly magnetized 
dense cores with dimensionless mass-to-flux ratios $\gtrsim 5$, but 
not in dense cores magnetized to higher, more typical levels. Our 
conclusion is that while the misalignment helps with disk formation, 
especially in relatively weakly magnetized cores, it may not 
provide a complete resolution to the magnetic braking catastrophe 
by itself.

The rest of the paper is organized as follows. In \S\ref{setup}, we
describe the model setup. The numerical results are described in
\S\ref{misalignment} and \S\ref{strong}. We compare our results to
those of \citeauthor{Joos+2012} and discuss their implications 
in \S\ref{discussion} and conclude with a short summary in \S\ref{summary}.

\section{Problem Setup}
\label{setup}

We follow \citet{Li+2011} and \citet{Krasnopolsky+2012} and start
our simulations from a uniform, spherical core of $1\msun$ and
radius $10^{17}\cm$ in a spherical coordinate system
$(r,\theta,\phi)$. The initial
density $\rho_0=4.77 \times 10^{-19}\gm\cm^{-3}$ corresponds to a
molecular hydrogen number density of $10^5\cm^{-3}$. We adopt an
isothermal equation of state with a
sound speed $a=0.2\kms$ below a critical density $\rho_c=10^{-13}
\gm\cm^{-3}$, and a polytropic equation of state $p\propto\rho^{5/3}$
above it. At the beginning of the simulation, we impose a solid-body
rotation of angular speed $\Omega_0=10^{-13}\second^{-1}$ on the core,
with axis along the north pole ($\theta=0$). It corresponds to a ratio
of rotational to gravitational binding energy of 0.025, which is
typical of the values inferred for NH$_3$ cores (\ct{Goodman+1993}).
The initial magnetic field is uniform, tilting away from the rotation
axis by an angle $\theta_0$. We consider three values for the
initial field: $B_0=10.6$, $21.3$ and $35.4\muG$, corresponding to
dimensionless mass-to-flux ratio, in units of $(2\pi G^{1/2})^{-1}$,
$\lambda = 9.72$, $4.86$ and $2.92$, respectively, for the core
as a whole. The mass-to-flux ratio for the central flux tube
$\lambda_c$ is higher than the global value $\lambda$ by $50\%$,
so that $\lambda_c= 14.6$, $7.29$ and $4.38$ for the three cases
respectively. The effective mass-to-flux ratio $\lambda_{\rm eff}$
should lie between these two limits.  If the star formation
efficiency per core is $\sim 1/3$ (e.g., \ct{Alves+2007}), then
one way to estimate $\lambda_{\rm eff}$ is to consider the (cylindrical)
magnetic flux surface that encloses $1/3$ of the core mass, which
yields $\lambda_{\rm eff}=1.41 \lambda$, corresponding to $13.7$,
$6.85$, and $4.12$ for the three cases respectively; the fraction
$1/3$ is also not far from the typical fraction of core mass that
has accreted onto the central object at the end of our simulations
(see Table~1). For the tilt angle, we also consider three values:
$\theta_0=0\degree$, $45\degree$ and $90\degree$. The $\theta_0=0\degree$
corresponds to the aligned case, with the magnetic field and
rotation axis both along the $z$-axis ($\theta=0$). The
$\theta_0=90\degree$ corresponds to the orthogonal case, with the
magnetic field along the $x$-axis ($\theta=90\degree$, $\phi=0$).
Models with these nine combinations of parameters are listed in
Table~1; additional models are discussed below.

\begin{deluxetable}{lllllll}
\tablecolumns{7}
\tablecaption{Models \label{table:first}}
\tablehead{
\colhead{Model}
& \colhead{$\lambda^a$}
& \colhead{$\lambda_{\rm eff}^b$}
& \colhead{$\theta_0$}
& \colhead{$\eta$ ($10^{17}\etaunit$)}
& \colhead{$M_*^c$ ($M_\odot$) }
& \colhead{RSD$^d$ }
}
\startdata
A   & 9.72  & 13.7 & 0$\degree$  & 1 & 0.24   & No \\ %(f461, ccc)
B   & 4.86  & 6.85 & 0$\degree$  & 1 &  0.22  & No \\  % (f490, ccc)
C   & 2.92  & 4.12 & 0$\degree$  &1 &   0.33  & No \\   % (f581, ccc)

D   & 9.72 & 13.7 & 45$\degree$  & 1 &  0.21  & Yes/Porous \\  % (f439, cccc)
E   & 4.86 & 6.85 & 45$\degree$ & 1 &  0.35  & No \\  % (f490, cccc)
F   & 2.92 & 4.12 & 45$\degree$  & 1 &  0.27   & No \\ % (f519, cont)

G   & 9.72 & 13.7 & 90$\degree$ & 1 &  0.38  & Yes/Robust   \\ %  (f533, ccccc)
H   & 4.86 & 6.85 & 90$\degree$  & 1 &  0.46  & Yes/Porous \\  % (f545, ccccc)
I   & 2.92 & 4.12 & 90$\degree$ & 1 &  0.47  & No \\ %  (f621, ccccc)

M   & 9.72 & 13.7 & 90$\degree$ &  0 & 0.10   & Yes/Robust \\
N   & 9.72 & 13.7 & 90$\degree$ &  0.1 & 0.26 & Yes/Robust \\ %  (f484, ccc,

P   & 4.03 & 5.68 & 90$\degree$ &  1 & 0.25   & Yes/Porous \\ % (f446, cccc, running)
Q   & 3.44 & 4.85 & 90$\degree$ &  1 & 0.14 & No \\ %  (f447, main, running)
\enddata
\tablecomments{a). The average dimensionless mass-to-flux ratio for
  the core as a whole; b). The effective mass-to-flux ratio for the
  central $1/3$ of the core mass (see \S\ref{setup} for discussion);
  c). Mass of the central object when the simulation is
  stopped; d). ``Robust'' disks are persistent, rotationally supported
  structures that rarely display large deviations from smooth
  Keplerian motions, whereas ``porous'' disks are highly active,
  rotationally dominated structures with large distortions and may
  occasionally be completely disrupted.}
\end{deluxetable}

As in \citet{Krasnopolsky+2012}, we choose a non-uniform grid
of $96\times 64\times 60$. In the radial direction, the inner
and outer boundaries are located at $r=10^{14}$ and $10^{17}\cm$,
respectively. The radial cell size is smallest near the inner
boundary ($5\times 10^{12}\cm$ or
$\sim 0.3\au$). It increases outward by a constant factor $\sim 1.08$
between adjacent cells. In the polar direction, we choose a
relatively large cell size ($7.5\degree$) near the polar axes,
to prevent the azimuthal cell size from becoming prohibitively
small; it decreases smoothly to a minimum of $\sim 0.63\degree$ near
the equator, where rotationally supported disks may form.
The grid is uniform in the azimuthal direction.

The boundary conditions in the azimuthal direction are periodic.
In the radial direction, we impose the standard outflow
boundary conditions. Material leaving the inner radial boundary
is collected as a point mass (protostar) at the center. It acts
on the matter in the computational domain through gravity. On
the polar axes, the boundary condition is chosen to be reflective.
Although this is not strictly valid, we expect its effect to be
limited to a small region near the axis.

We initially intended to carry out simulations in the
ideal MHD limit, so that they can be compared more
directly with other work, especially \citet{Joos+2012}.
However, ideal MHD simulations tend to produce numerical ``hot zones''
that force the calculation to stop early in the protostellar
mass accretion phase, a tendency we noted in our previous 2D
(\ct{MellonLi2008}) and 3D simulations (\ct{Krasnopolsky+2012}).
To lengthen the simulation, we include a small, spatially
uniform resistivity $\eta=10^{17}\etaunit$. We have
verified that, in the particular case Model G ($\lambda=9.72$
and $\theta_0=90\degree$), this resistivity changes the flow
structure little compared to either the ideal MHD Model M
(before the latter stops) or Model N, where the resistivity
is reduced by a factor 10, to $10^{16}\etaunit$.

\section{Weak-Field Case: Disk Formation Enabled by Field-Rotation
Misalignment}
\label{misalignment}

\subsection{Equatorial Pseudodisk vs Magnetically Induced Curtain}

\begin{figure}
\epsscale{1.2}
\plottwo{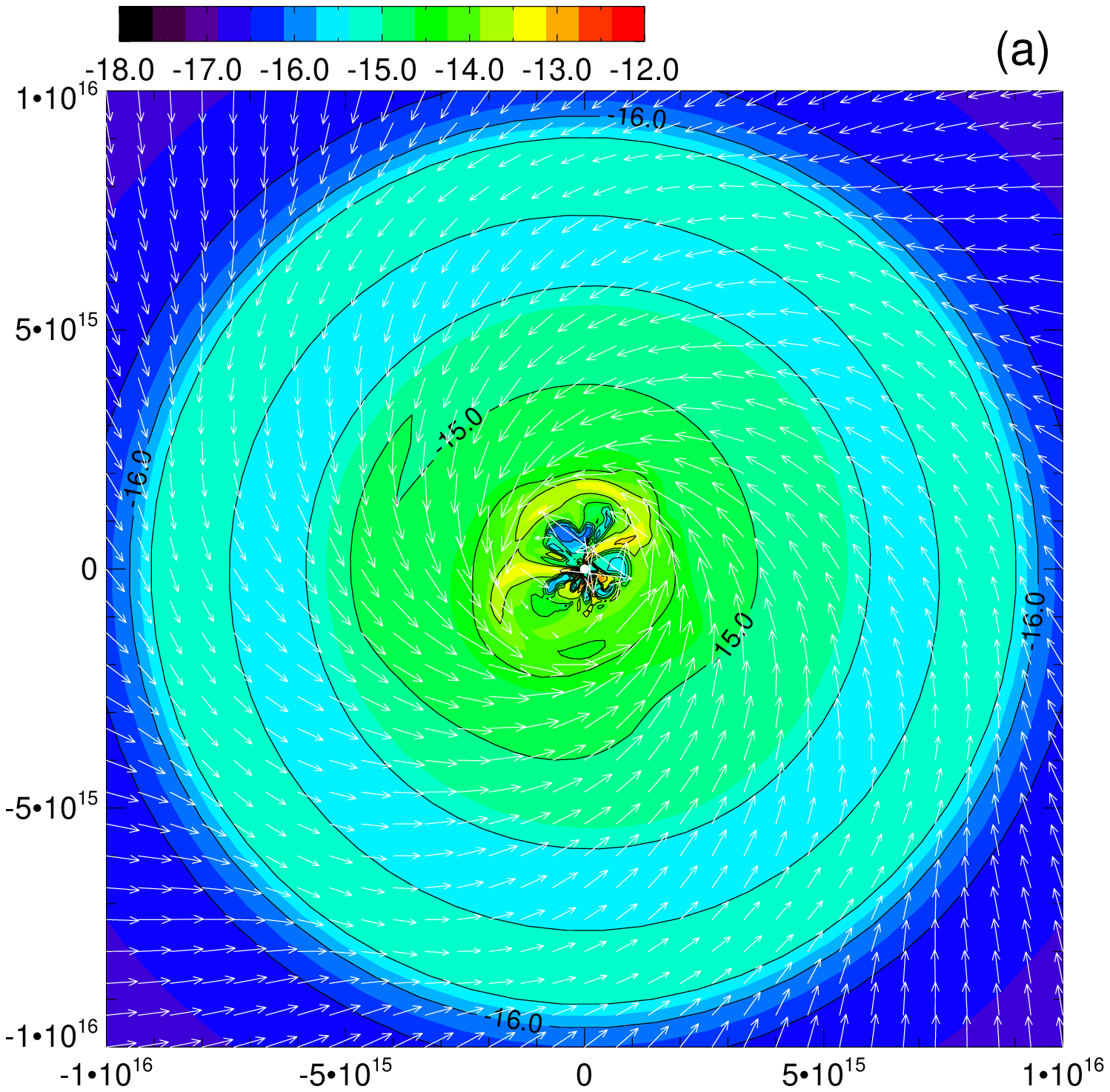}{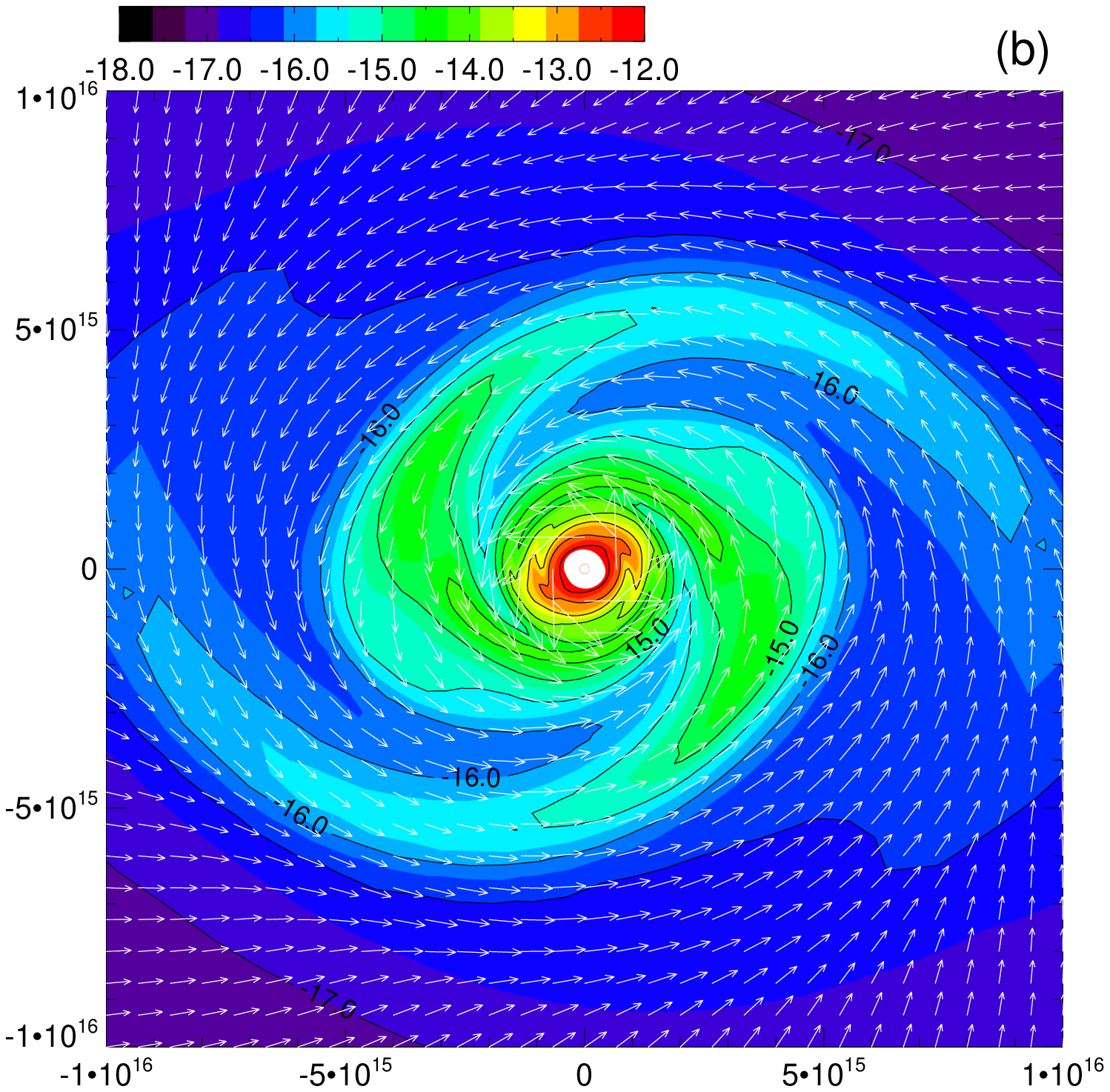}
\plottwo{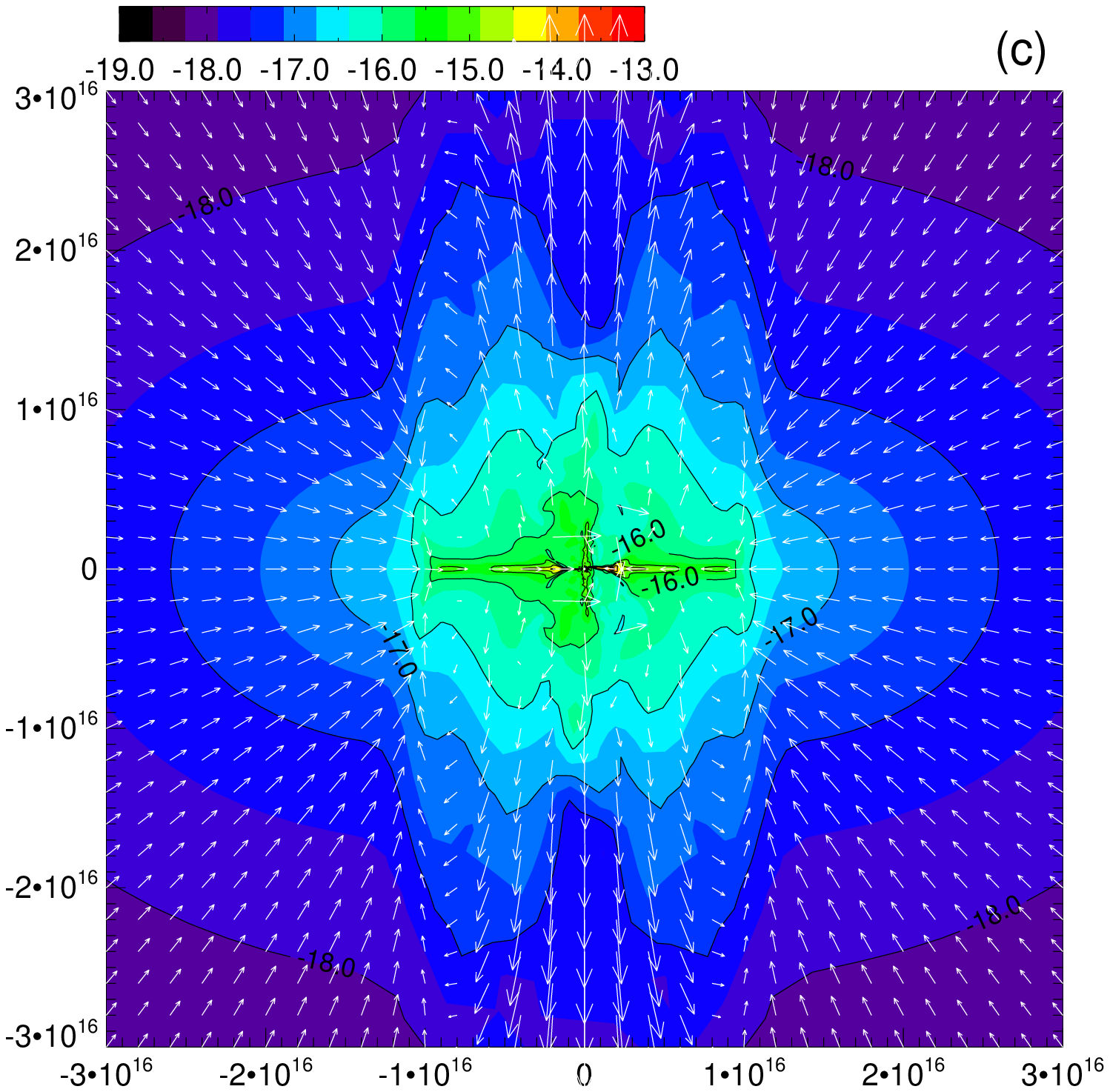}{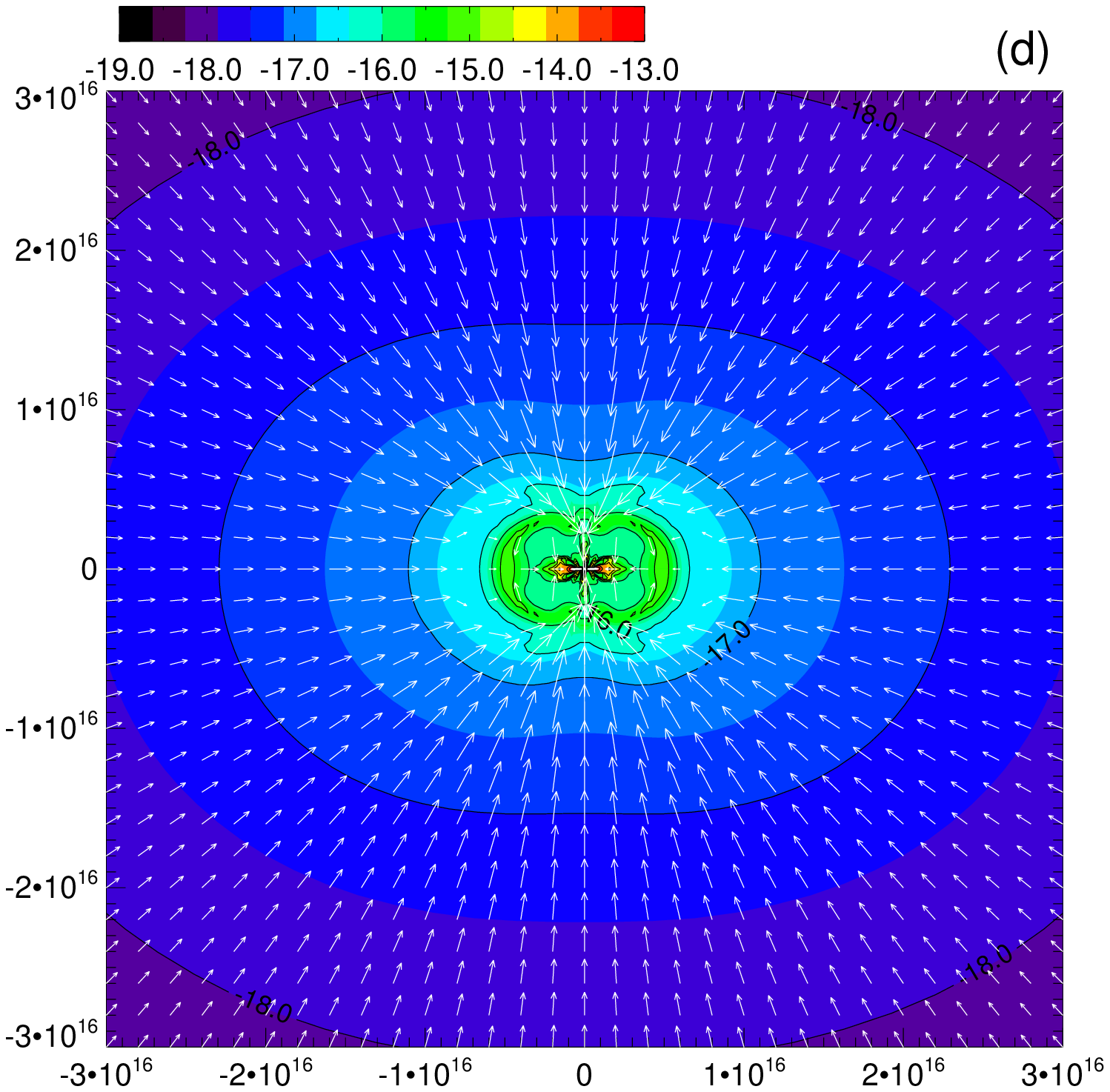}

\caption{Snapshots of the logarithm of density (color map, in units of
  $\gram\cm^{-3}$) and velocity field (white arrows) for the weakly
  magnetized core of $\lambda=9.72$ at a representative time
$t=3.9\times 10^{12}\second$. The left (right) panels are for the aligned
(orthogonal) case, and the top (bottom) panels are for the equatorial
$x$-$y$ (meridian $y$-$z$) plane. Note the powerful bipolar outflow in
  panel (c) driven by the pseudodisk in panel (a). The lack of
  powerful outflow in panel (d) is indicative of a weaker magnetic
  braking, which is consistent with the presence of a rotationally
supported disk in panel (b). The length unit is cm. The scale in
(c) and (d) is three times larger than that in (a) and (b). }
\label{contrast}
\end{figure}

To illustrate the effect of the misalignment between the magnetic
field direction and rotation axis, we first consider an extreme case
where the magnetic field is rather weak (with a mass-to-flux ratio
$\lambda=9.72$ for the core as a whole and $\lambda_{\rm eff}=13.7$
for the inner $1/3$ of the core mass). In this case, a well-formed
rotationally supported disk is present in the orthogonal case with
$\theta_0=90\degree$ (Model G in Table~1). Such a disk is absent in
 the aligned case (with $\theta_0=0\degree$, Model A). The
contrast is illustrated in
Fig.\ \ref{contrast}, where we plot snapshots of the aligned and
orthogonal cases at a representative time $t=3.9\times 10^{12}\second$,
when a central mass of $0.11$ and $0.12\msun$, respectively, has
formed. The flow structures in the two cases are very different in
both the equatorial (panels [a] and [b]) and meridian (panels [c]
and [d]) plane. In the equatorial plane, the aligned case has a
relatively large (with radius
$\sim 10^{16}\cm$) over-dense region where material spirals rapidly
inward. On the (smaller) scale of $10^{15}\cm$, the structure is
dominated by expanding, low-density lobes; they are the decoupling
enabled magnetic structures (DEMS for short) that have been
studied in detail by \citet{Zhao+2011} and \citet{Krasnopolsky+2012}.
No rotationally supported disk is evident. The equatorial
structure on the $10^{16}\cm$ scale in the orthogonal case is
dominated by a pair of spirals instead.
The spirals merge, on the $10^{15}\cm$ scale, into a more or less
continuous, rapidly rotating structure --- a rotationally supported
disk. Clearly, the accretion flow in the orthogonal case was able
to retain more angular momentum than in the aligned case. Why is
this the case?

A clue comes from the meridian view of the two cases (panels [c] and
[d] of Fig.\ \ref{contrast}). In the aligned case, there is a strong
bipolar outflow extending beyond $3\times 10^{16}\cm$ at the
relatively early time shown. The outflow forces most of the infalling
material to accrete through a flattened equatorial structure --- an
over-dense pseudodisk (\ct{GalliShu1993}; see panel [a] for a face-on
view of the pseudodisk, noting the difference in scale between
panel [c] and [a]). It is the winding of the magnetic field lines
by the rotating material in the pseudodisk that drives the bipolar
outflow in the first place. The wound-up field lines act back on the
pseudodisk material, braking its rotation. It is the efficient
magnetic braking in the pseudodisk that makes it difficult for
rotationally supported disks to form in the aligned case.

\begin{figure}
\epsscale{0.6}
\plotone{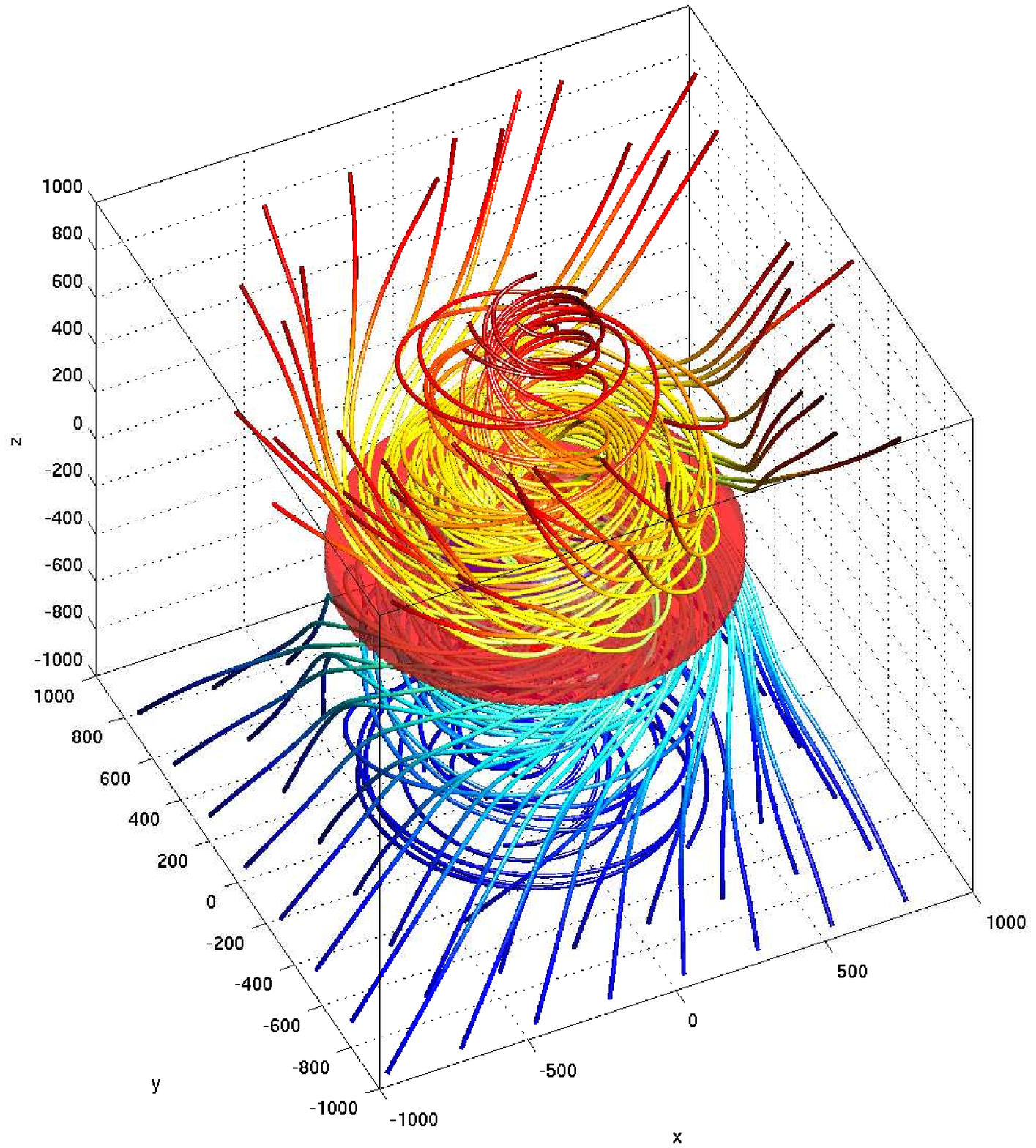}
\plotone{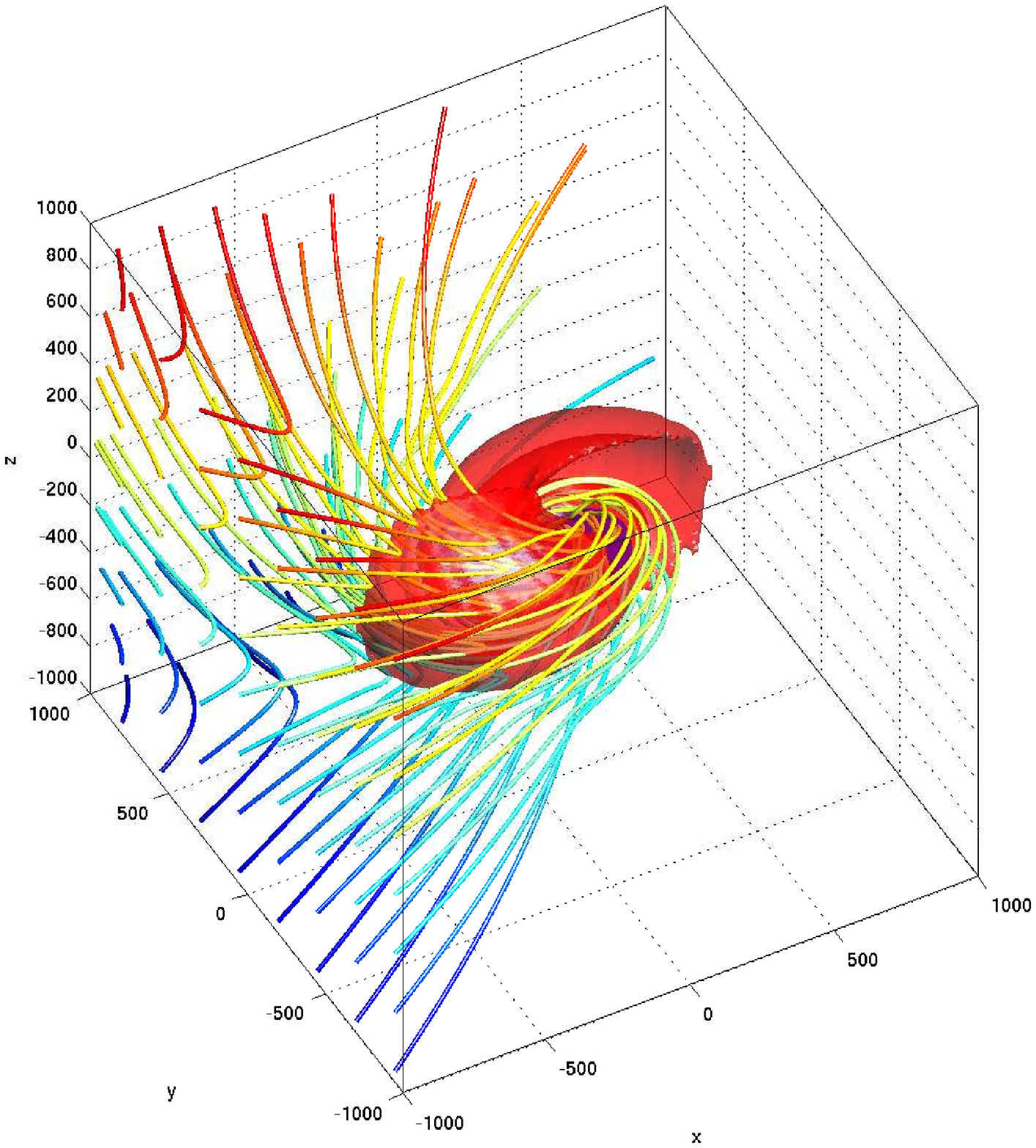}

\caption{3D view of representative magnetic field lines and isodensity
  surfaces at $\rho=10^{-16}$ (red) and $10^{-15}\gram\cm^{-3}$ (blue)
  for the aligned Model A and orthogonal Model G. For clarity, only
  field lines
  originated from the bottom $x$-$y$ and left $y$-$z$ plane are plotted,
  respectively. Note that the magnetically induced equatorial
  pseudodisk in the aligned case is warped by rotation into two
  snail-shaped curtains that spiral inward to form a disk (blue
  surface) in the orthogonal case. The length is in units of AU.}
\label{3D}
\end{figure}

The prominent bipolar outflow indicative of efficient magnetic braking
is absent in the orthogonal case, as was emphasized by
\citet{CiardiHennebelle2010}. It is replaced by a much smaller,
shell-like structure inside which the $10^{15}\cm$-scale rotationally
supported disk is encased (panel [d]). To understand this difference
in flow structure pictorially, we plot in Fig.\ \ref{3D} the
three-dimensional structure of the magnetic field lines on the
scale of $1000\AU$ (or $1.5\times 10^{16}\cm$), which is $50\%$ larger
than the size of panels (a) and (b) of Fig.\ \ref{contrast}, but half
of that of panels (c) and (d). Clearly, in the aligned case, the
relatively weak initial magnetic field (corresponding to $\lambda
\approx 10$) has been wound up many turns by the material in the
equatorial pseudodisk, building up a magnetic pressure in the
equatorial region that is released along the polar directions
(see the first panel of Fig.\ \ref{3D}). The magnetic pressure
gradient drives a bipolar outflow, which is evident in panel
(c) and in many previous simulations of magnetized core collapse,
including the early ones such as \citet{Tomisaka1998} and \citet{Allen+2003}.
In contrast, in the orthogonal case, the equatorial
region is no longer the region of the magnetically induced
pseudodisk. In the absence of rotation (along the $z$-axis), the
dense core material would preferentially contract along the
field lines (that are initially along the $x$-axis) to form a
dense sheet in the $y$-$z$ plane that passes through the
origin. The twisting of this sheet by rotation along
the $z$-axis produces two curved ``curtains'' that
spiral smoothly into the disk at small distances, somewhat
analogous in shape to two snail-shells (see the second panel
of Fig.\ \ref{3D}).

\begin{figure}
\epsscale{0.75}
\plotone{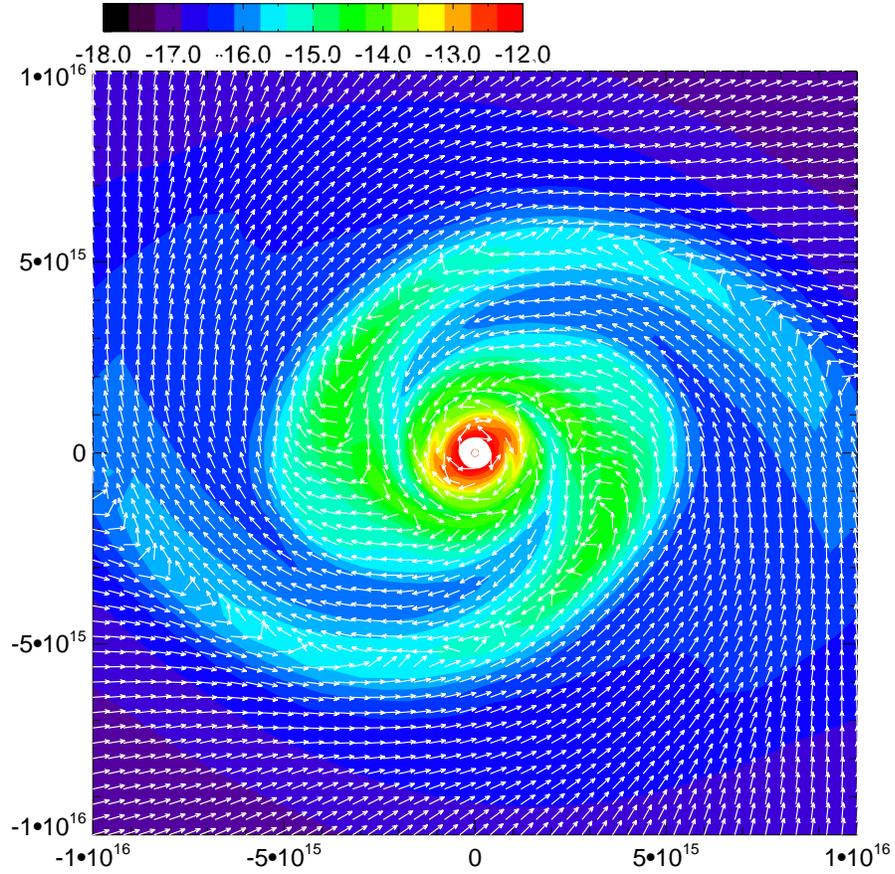}
\caption{Same as panel (b) of Fig.\ \ref{contrast} but with magnetic
  unit vectors instead of velocity vectors plotted. Note that the
  field lines kink sharply in the spirals. The kinks demonstrate
  that the spirals (and the curtain that contains them) are the
  counterparts to the equatorial pseudodisk in the aligned case.}
\label{pinch}
\end{figure}

The snail-shaped dense curtain in the orthogonal case naturally
explains the morphology of the density maps shown in panels (b) and
(d) of Fig.\ \ref{contrast}. First, the two prominent spiral arms in
the panel (a) are simply the equatorial ($x$-$y$) cut of the curved
curtains. An interesting feature of the spirals (and the snail-shaped
dense curtain as a whole) is that they are the
region where the magnetic field lines change directions sharply. This
is illustrated in Fig.\ \ref{pinch}, which is similar to panel (b) of
Fig.\ \ref{contrast}, except that the magnetic vectors (rather than
velocity vectors) are plotted on top of the density map.
Clearly, the spirals separate the field lines rotating counter clock-wise
(lower-right part of the figure) from those rotating close-wise
(upper-left). The sharp kink is analogous to the well-known field line
kink across the equatorial pseudodisk in the aligned case, where the
radial component of the magnetic field changes direction. It supports
our interpretation of the spirals and, by extension, the
curtain as a magnetically induced feature, as is the case of
pseudodisk. In other words, the spirals are not produced by
gravitational instability in a rotationally supported structure; they
are ``pseudospirals'' in the same sense as the ``pseudodisks'' of
\citet{GalliShu1993}. The field line kinks are also evident across the
dense curtain in the 3D structure shown in the second panel of
Fig. \ref{3D}.

The 3D topology of the magnetic field and the dense structures that it
induces lie at the heart of the difference in the magnetic braking
efficiency between the aligned and orthogonal case. In particular, a
flattened, rotating, equatorial pseudodisk threaded by an ordered
magnetic field with an appreciable vertical component (along the
rotation axis) is more conducive to driving an outflow than a warped
curtain with a magnetic field predominantly tangential to its
surface.
The outflow plays a key role in angular momentum removal and the
suppression of rotationally supported disks, as we demonstrate
next.

\subsection{Torque Analysis}
\label{torque}

To quantify the outflow effect, we follow
\citeauthor{ZhaoLi2013} (\citeyear{ZhaoLi2013}; see also \ct{Joos+2012}) and
compare the rates of angular momentum change inside a finite volume
$V$ through its surface $S$ due to infall and outflow to that due to
magnetic torque. The total magnetic torque relative to the origin
(from which a radius vector ${\bf r}$ is defined) is
\begin{equation}
{\bf N}_{m}=\frac{1}{4\pi} \int[{\bf r} \times ((\nabla \times {\bf
    B}) \times {\bf B})]\,dV,
\end{equation}
where the integration is over the volume $V$. Typically, the magnetic
torque comes mainly from the magnetic tension rather than pressure
force. The dominant magnetic tension term can be simplified to a
surface integral (\ct{MatsumotoTomisaka2004})
\begin{equation}
{\bf N}_t=\frac{1}{4\pi} \int ({\bf r} \times {\bf B})({\bf B} \cdot d{\bf S}),
\end{equation}
over the surface $S$ of the volume. This volume-integrated magnetic
torque is to be compared with the rate of angular momentum advected
into the volume through fluid motion,
\begin{equation}
{\bf N}_a=-\int \rho({\bf r} \times {\bf v})({\bf v} \cdot d{\bf S}),
\end{equation}
which will be referred to as the advective torque below.

Since the initial angular momentum of the dense core is along the
$z$-axis, we will be mainly concerned with the $z$-component of
the magnetic and advective torque which, for a spherical volume
inside radius $r$, are given by
\begin{equation}
N_{t,z} = \frac{1}{4\pi} \int \varpi B_\phi B_r dS,
\end{equation}
and
\begin{equation}
N_{a,z} = -\int \rho \varpi v_\phi v_r dS.
\end{equation}
The advective torque consists of two parts: the rates of angular
momentum advected into and out of the sphere by infall and
outflow respectively:
\begin{equation}
N_{a,z}^{\rm{in}} = -\int \rho \varpi v_\phi v_r (< 0) dS,
\end{equation}
and
\begin{equation}
N_{a,z}^{\rm{out}} = -\int \rho \varpi v_\phi v_r (> 0) dS.
\end{equation}

\begin{figure}
\epsscale{1.0}
\plottwo{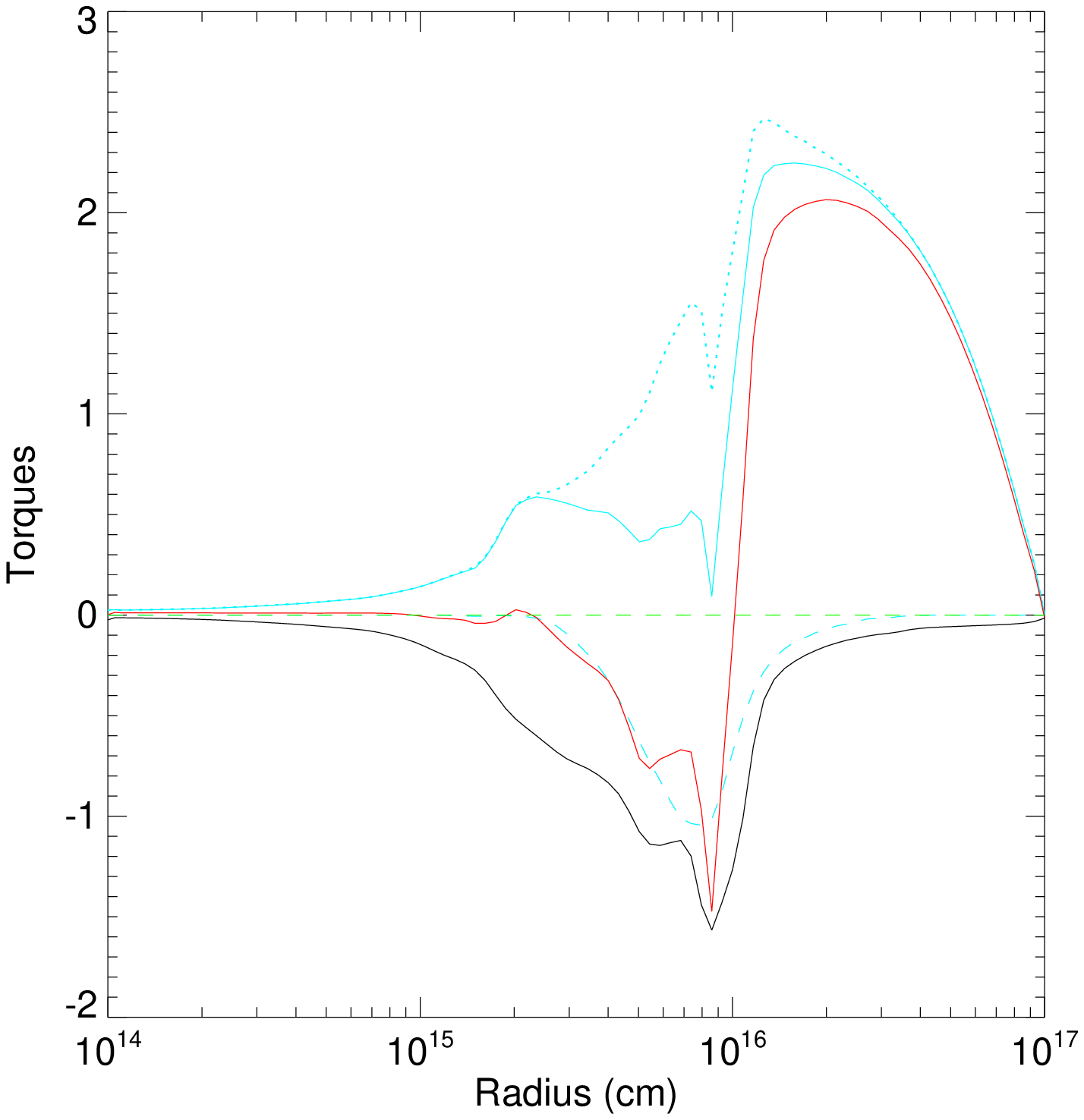}{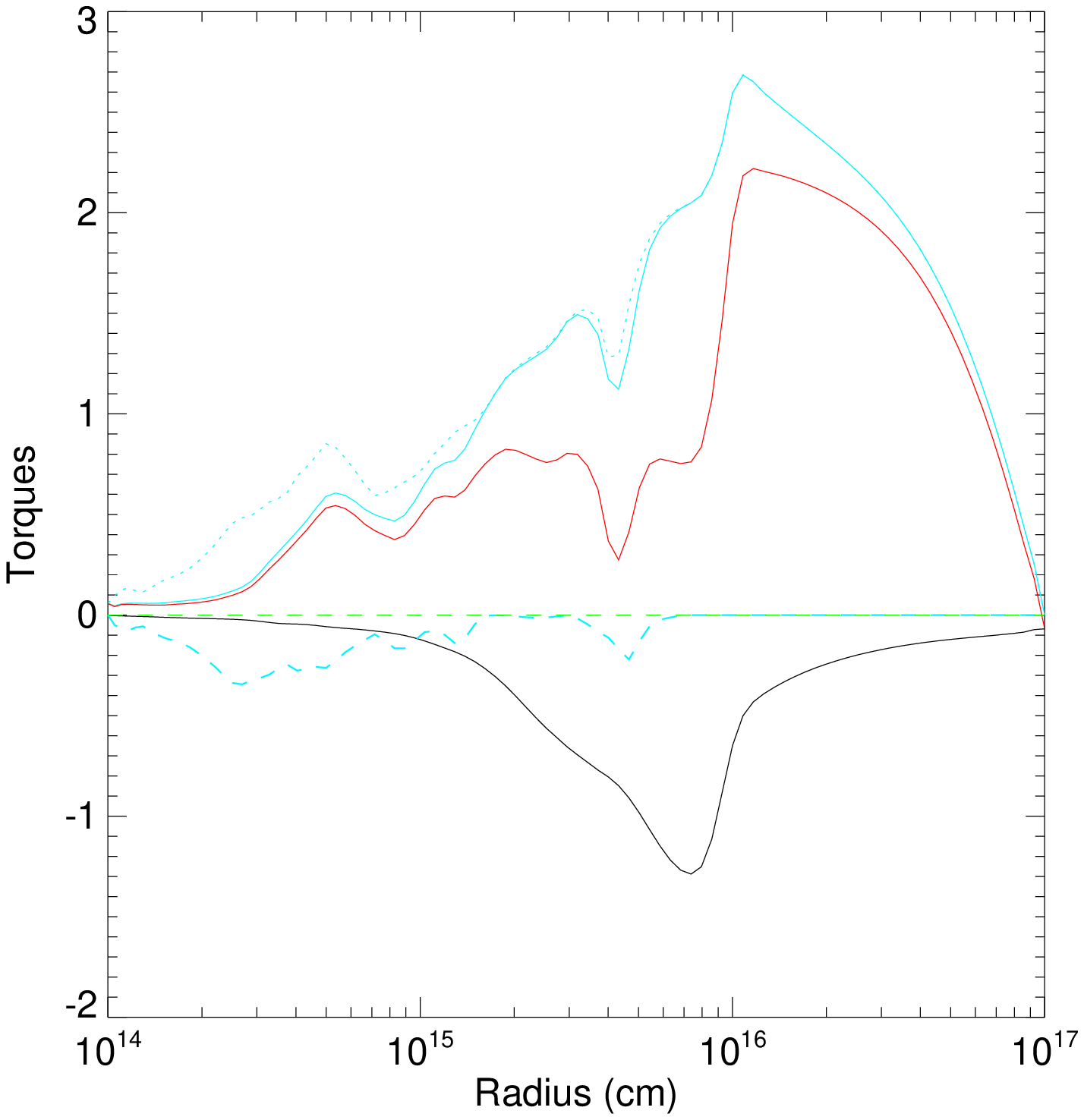}
\caption{Magnetic (black solid line, $N_{t,z}$) and total advective
  (magenta, $N_{a,z}$) torque acting on spheres of different radii
  for the aligned (left panel) and orthogonal (right) case at the
  representative time $t=3.9\times 10^{12}\second$. Also
  plotted are the net torque (red, $N_{t,z}+N_{a,z}$), and the
  contributions to the advective torque by infall (dotted,
  $N_{a,z}^{\rm{in}}$) and outflow (dashed, $N_{a,z}^{\rm{out}}$). Note that the
net torque is negative between $\sim 2\times 10^{15}$ and $\sim
10^{16}\cm$ for the aligned case but positive for the orthogonal case,
mainly because the outflow removes more angular momentum in the former
than in the latter.}
\label{Torque}
\end{figure}

An example of the magnetic and advective torques is shown in
Fig.\ \ref{Torque}. The torques are evaluated on spherical
surfaces of different radii, at the representative time
$t=3.9\times 10^{12}\second$. For the aligned case, the net torque
close to the central object is nearly zero up to a radius
of $\sim 2\times 10^{15}\cm$, indicating that the angular
momentum advected inward is nearly completely removed by magnetic
braking there. At larger distances, between $\sim 2\times 10^{15}$
and $\sim 10^{16}\cm$, the net torque  $N_{t,z}+N_{a,z}$ is negative,
indicating that the angular momentum of the material inside a
sphere of radius in this range decreases with time. This is in
sharp contrast with the orthogonal case, where the net torque
is positive in that radial range, with the angular momentum
there increasing (rather than decreasing) with time. One may
think that the difference is mainly due to a significantly larger
magnetic torque $N_{t,z}$ in the aligned case than in the orthogonal
case. Although this is typically the case at early times, the
magnetic torques in the two cases become comparable at later
times (see the lowest solid lines in
the two panels of Fig.\ \ref{Torque}; a movie of the torques
is available on request from the authors). A bigger difference
comes from the total (or net) angular momentum $N_{a,z}=N_{a,z}^{\rm{in}}
+N_{a,z}^{\rm{out}}$ advected inward, which is substantially smaller
in the aligned case than in the orthogonal case (see the
uppermost solid lines in the two panels). The main reason for
the difference is that a good fraction of the angular momentum
advected inward by infall $N_{a,z}^{\rm{in}}$ is advected back out
by outflow $N_{a,z}^{\rm{out}}$ in the former, but not the latter.
This is helped by the fact that $N_{a,z}^{\rm{in}}$ is somewhat
smaller in the aligned case to begin with (compare the dotted
lines in the two panels). The lack of appreciable outward
advection of angular momentum by outflow, which is itself a
product of field-winding and magnetic braking in the aligned
case, appears to be the main reason for the orthogonal case
to retain more angular momentum at small radii and form a
rotationally supported disk in this particular case of
relatively weak magnetic field.

\begin{figure}
\epsscale{0.30}
\plotone{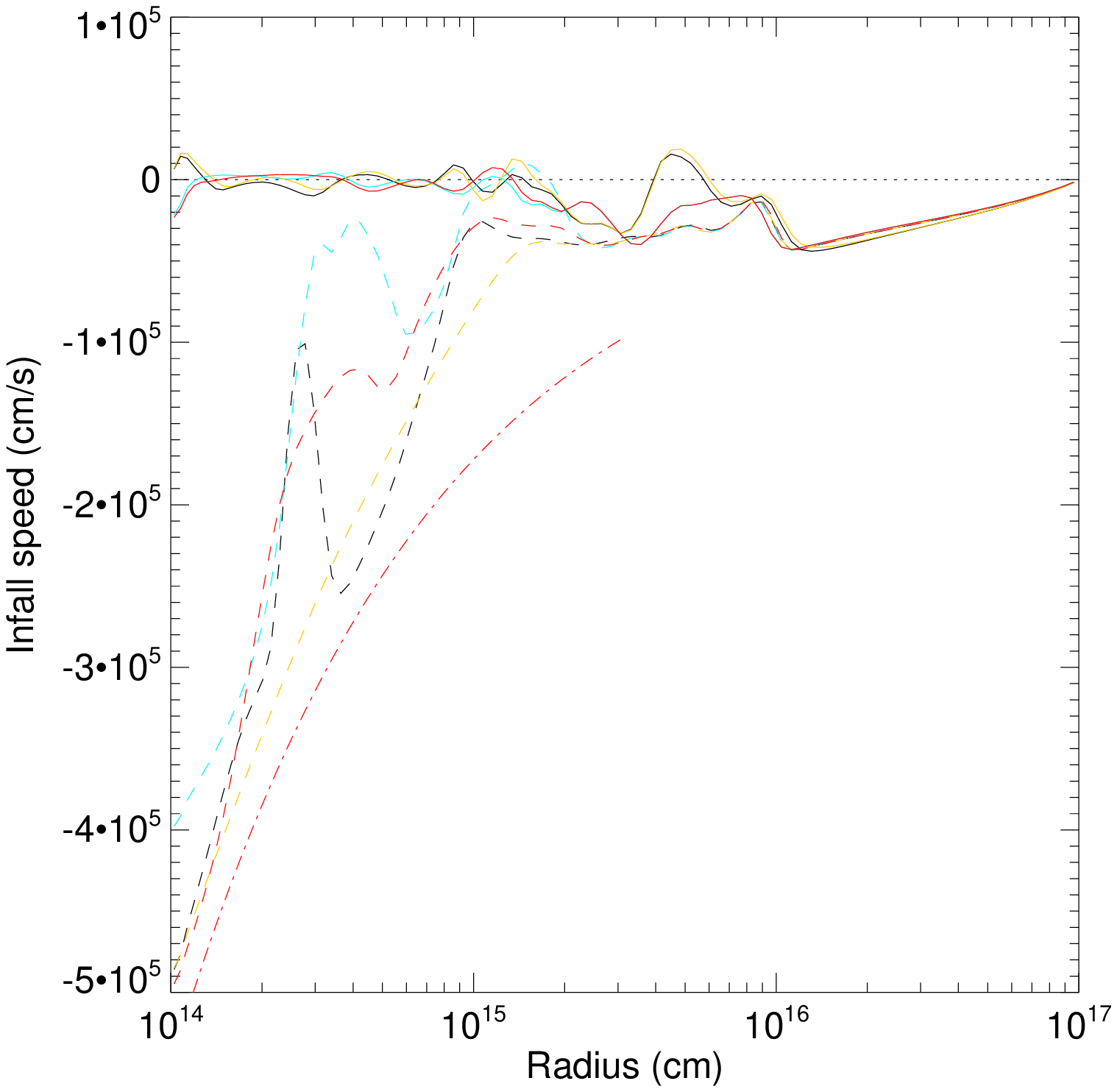}
\plotone{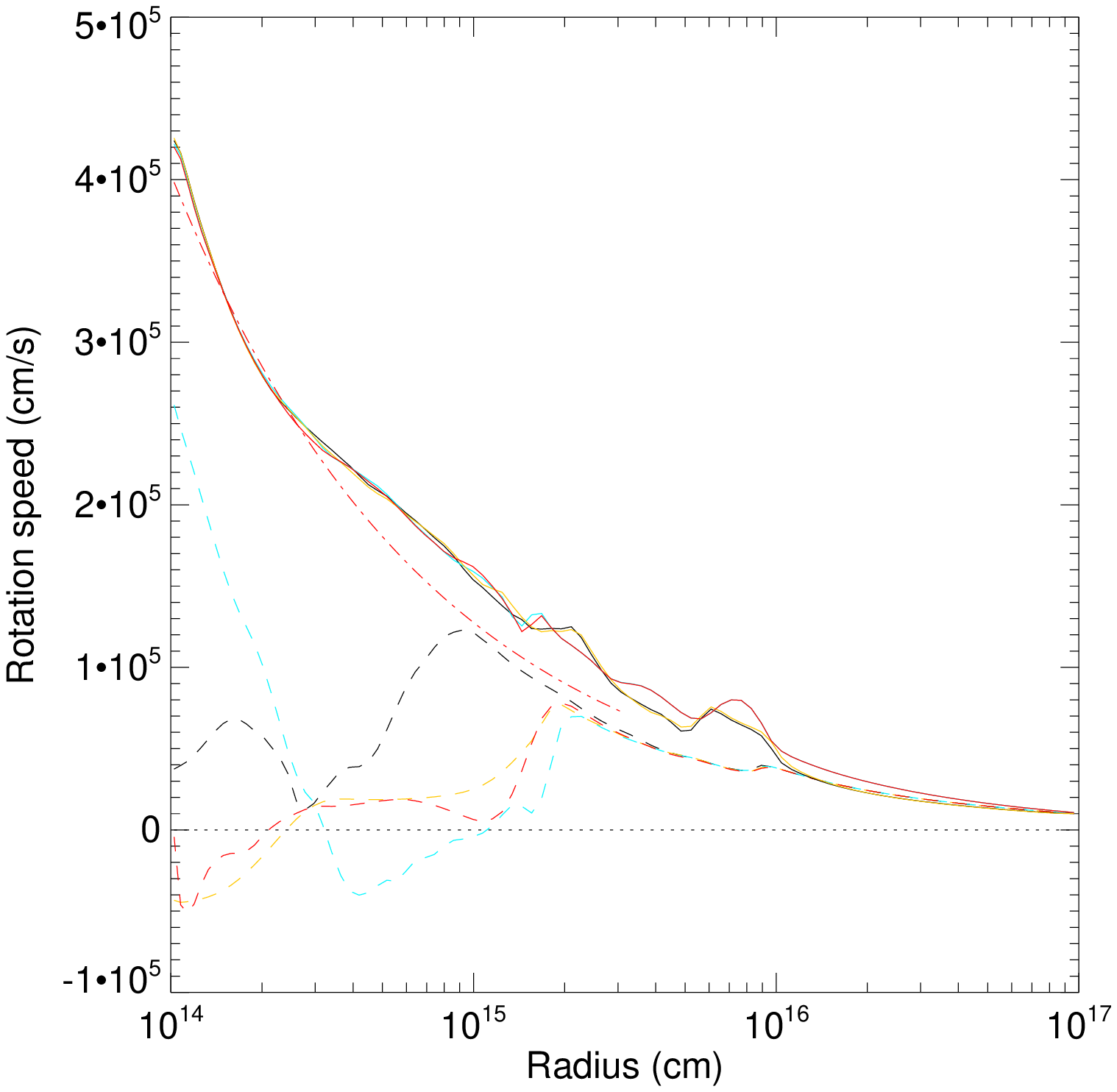}
\plotone{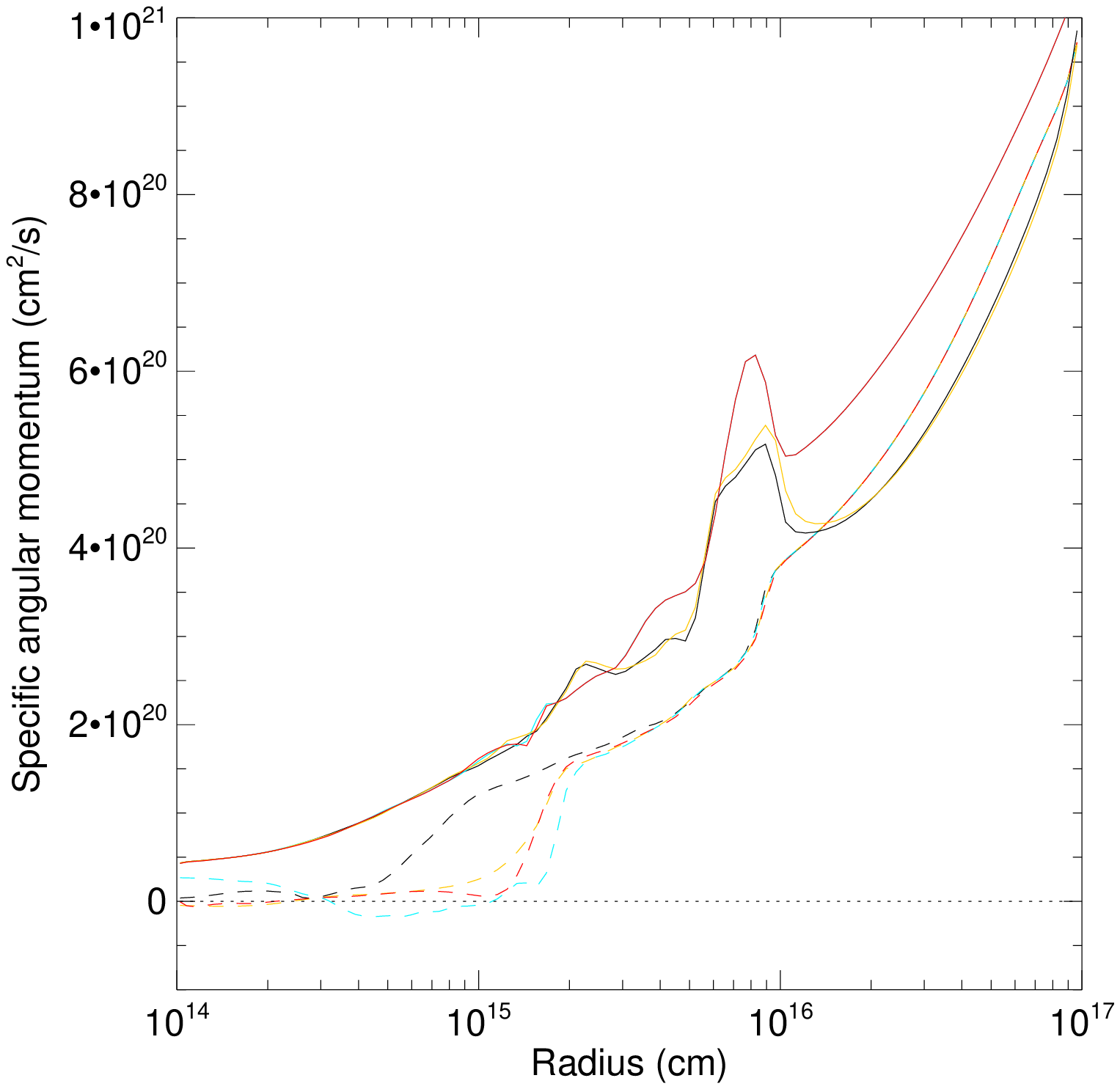}
\caption{Infall (left panel) and rotation (middle) speed and the
  specific angular momentum (right) on the equatorial plane along
  4 representative directions for the aligned (dashed lines) and
  orthogonal (solid) case. For comparison, the free-fall speed for
  the aligned case and the Keplerian profile for the orthogonal
  case are plotted as dotted-dash line in the left and
middle panel respectively.
  It is evident that a rotationally supported
  disk is formed in the orthogonal but not the aligned case. In the
  latter case, the specific angular momentum of the equatorial flow
  decreases significantly upon entering the pseudodisk at $\sim
  10^{16}\cm$ and the DEMS-dominated region around $\sim 10^{15}\cm$.}
\label{Rotation}
\end{figure}

The formation of a rotationally supported disk can be seen most
clearly in Fig.\ \ref{Rotation}, where we plot the infall and
rotation speed, as well as specific angular momentum as a
function of radius along 4 ($\pm x$ and $\pm y$) directions in the
equatorial plane. In the orthogonal case, the infall and rotation
speeds display the two tell-tale signs of rotationally supported
disks: (1) a slow, subsonic (although nonzero) infall speed much smaller
than the free fall value, and (2) a much faster rotation speed close to
the Keplerian value inside a radius of $\sim 100\AU$. The absence of
a rotationally supported disk in the aligned case is just as obvious.
It has a rotation speed well below the Keplerian value and an infall
speed close to the free fall value, especially at small radii up to
$\sim 100\AU$. This corresponds to the region dominated by the
low-density, strongly magnetized, expanding lobes (i.e., DEMS; see
panel [a] of Fig\ \ref{contrast}) where the angular momentum is almost
completely removed by a combination of magnetic torque and outflow
(see the third panel in Fig.\ \ref{Rotation}). Also evident from the
panel is that the specific angular momentum of the equatorial inflow
drops significantly twice: near $\sim 10^{15}$ and $\sim 10^{16}\cm$
respectively. The former corresponds to the DEMS-dominated region,
and the latter the pseudodisk (see panel [a] of Fig.\ \ref{contrast}).
The relatively slow infall inside the pseudodisk allows more time
for magnetic braking to remove angular momentum. It is the pseudodisk
(and its associated outflow) working in tandem with the DEMS that
suppresses the formation of a rotationally supported disk in the
aligned case. Interestingly, there is a bump near $\sim 10^{16}\cm$
for the specific angular momentum of the orthogonal case, indicating
that the angular momentum in the equatorial plane is transported
radially outward along the spiraling field lines from small to large
distances (see Fig.\ \ref{pinch}).

\begin{figure}
\epsscale{1.0}
\plottwo{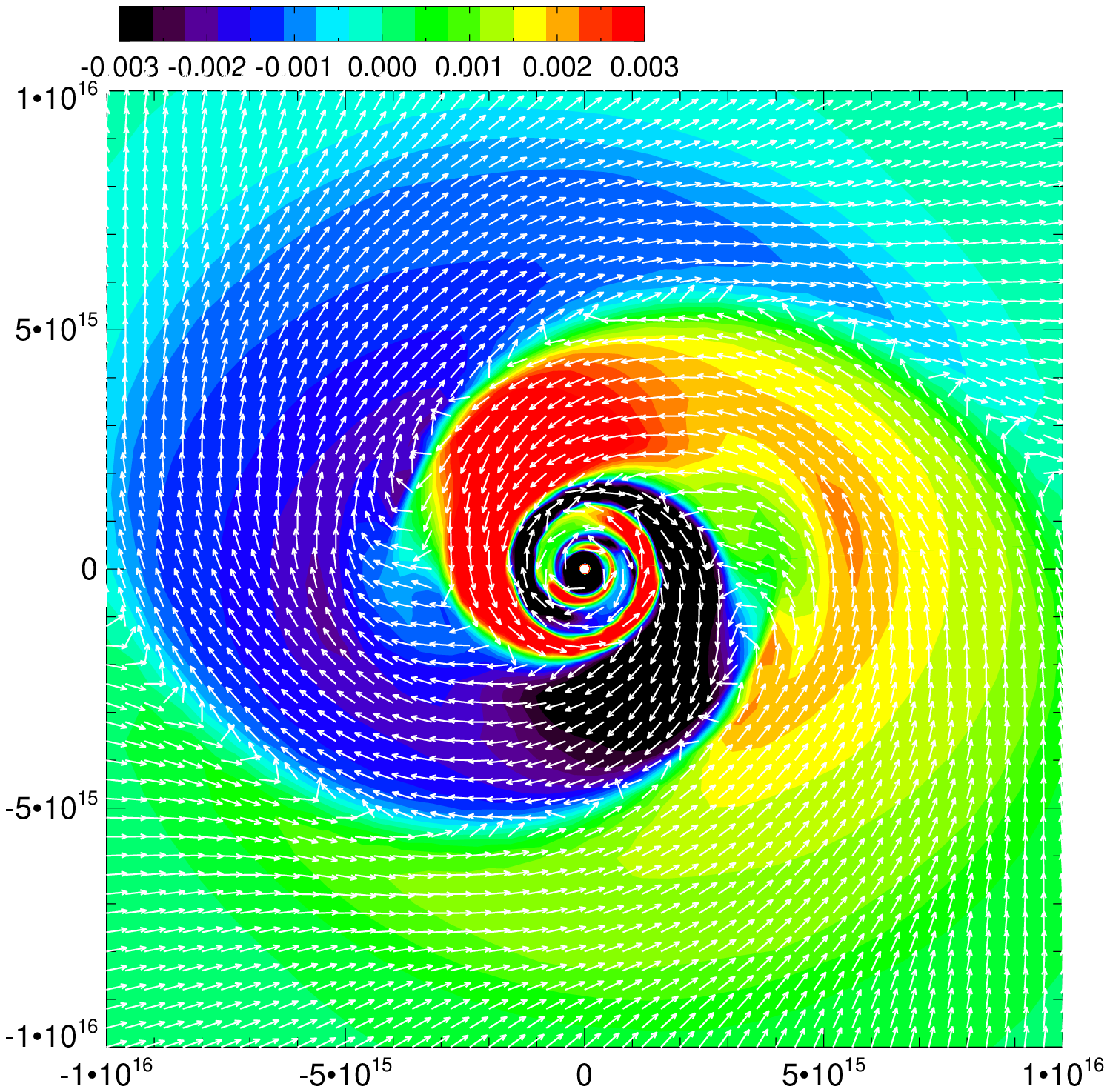}{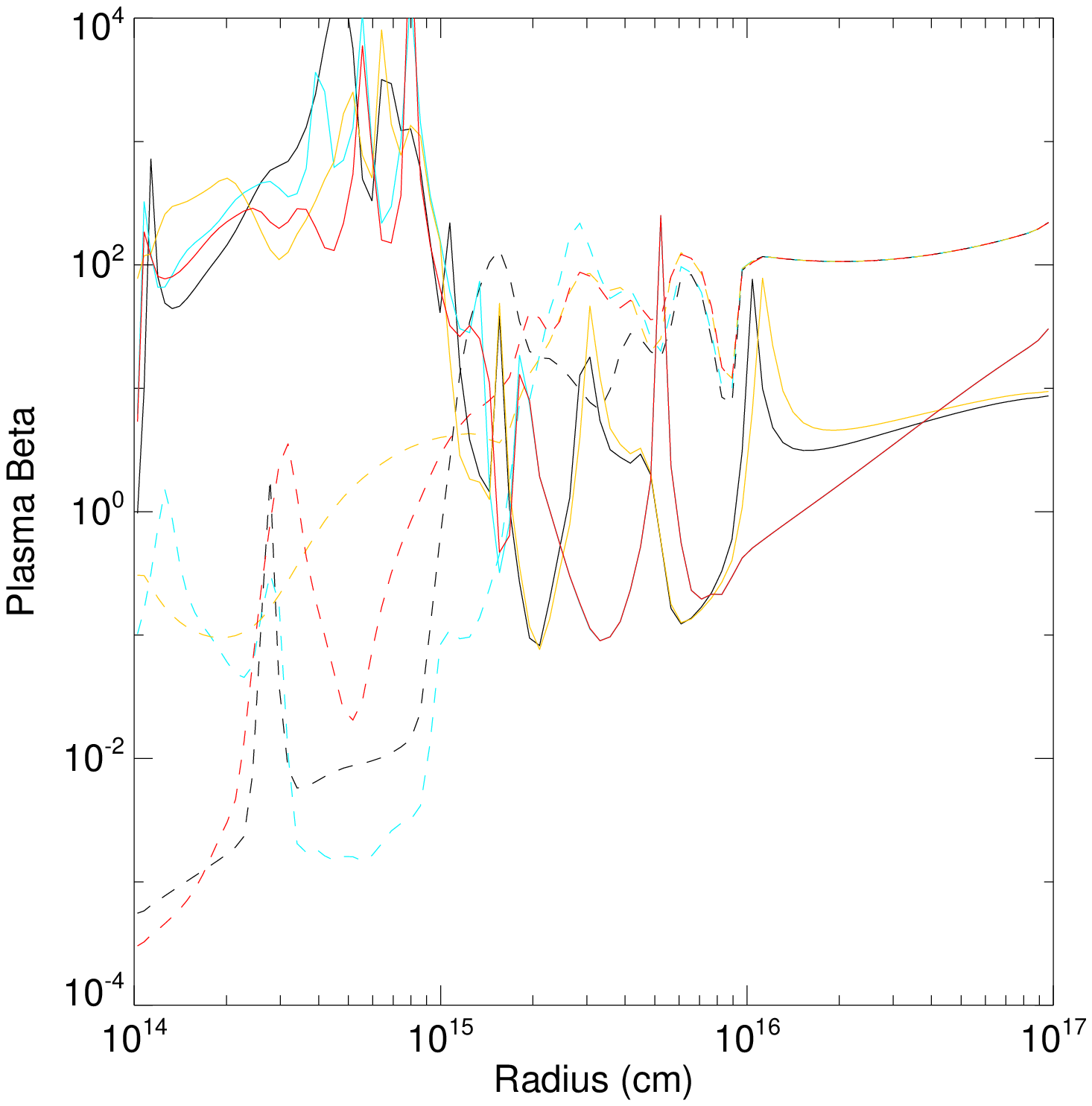}

\caption{Left: Same as Fig.\ \ref{pinch}, except that the color map is
  for the toroidal component of the equatorial magnetic field (rather
  than density), showing the inward spiraling of two strains of
  magnetic field lines of opposite polarity. For clarify, the plotted
  field strength is capped from above and below at
  $\pm 3\times 10^{-3}{\rm G}$.
  Right: The plasma $\beta$ along 4 representative
  directions in the equatorial plane showing that the inner $100\AU$
  region in the orthogonal case (i.e., on the Keplerian disk; solid
  lines) is dominated by thermal rather than magnetic pressure,
  whereas the opposite is true for the aligned case (dashed).}
\label{magnetic}
\end{figure}

The spiraling equatorial field lines in the orthogonal Model G have
an interesting property: they consist of two strains of opposite
polarity. As the strains get wound up more and more tightly by
rotation at smaller and smaller radii, field lines of opposite
polarity are pressed closer and closer together, creating a
situation that is conducive to reconnection, either of physical or
numerical origin (see the first panel of Fig.\ \ref{magnetic}).
Model G contains a small but finite resistivity
($\eta=10^{17}\etaunit$). It does not appear to
be responsible for the formation and survival of the Keplerian disk,
because a similar disk is also formed at the same (relatively
early) time for a smaller resistivity of $\eta=10^{16}\etaunit$
(Model N) and even without any explicit resistivity (Model M).
Numerical resistivity may have played a role here, but it is
difficult to quantify at the moment. In any case, the magnetic
field on the Keplerian disk appears to be rather weak, as can
be seen from the second panel of Fig.\ \ref{magnetic}, where the
plasma $\beta$ is plotted along 4 ($\pm x$ and $\pm y$) directions
in the equatorial plane. On the Keplerian disk in the orthogonal
Model G (inside $\sim 100\AU$), $\beta$ is of order $10^2$ or
more, indicating that there is more matter accumulating in the
disk than magnetic field, either because the matter slides
along the field lines into the disk (increasing density but
not the field strength) or because of numerical reconnection
that weakens the field, or both. This situation is drastically
different from the aligned case, where the inner $100\AU$ region
is heavily influenced by the magnetically-dominated low-density
lobes.

\section{Moderately Strong Field Cases: Difficulty with Disk Formation}
\label{strong}

\begin{figure}
\epsscale{1.0}
\plotone{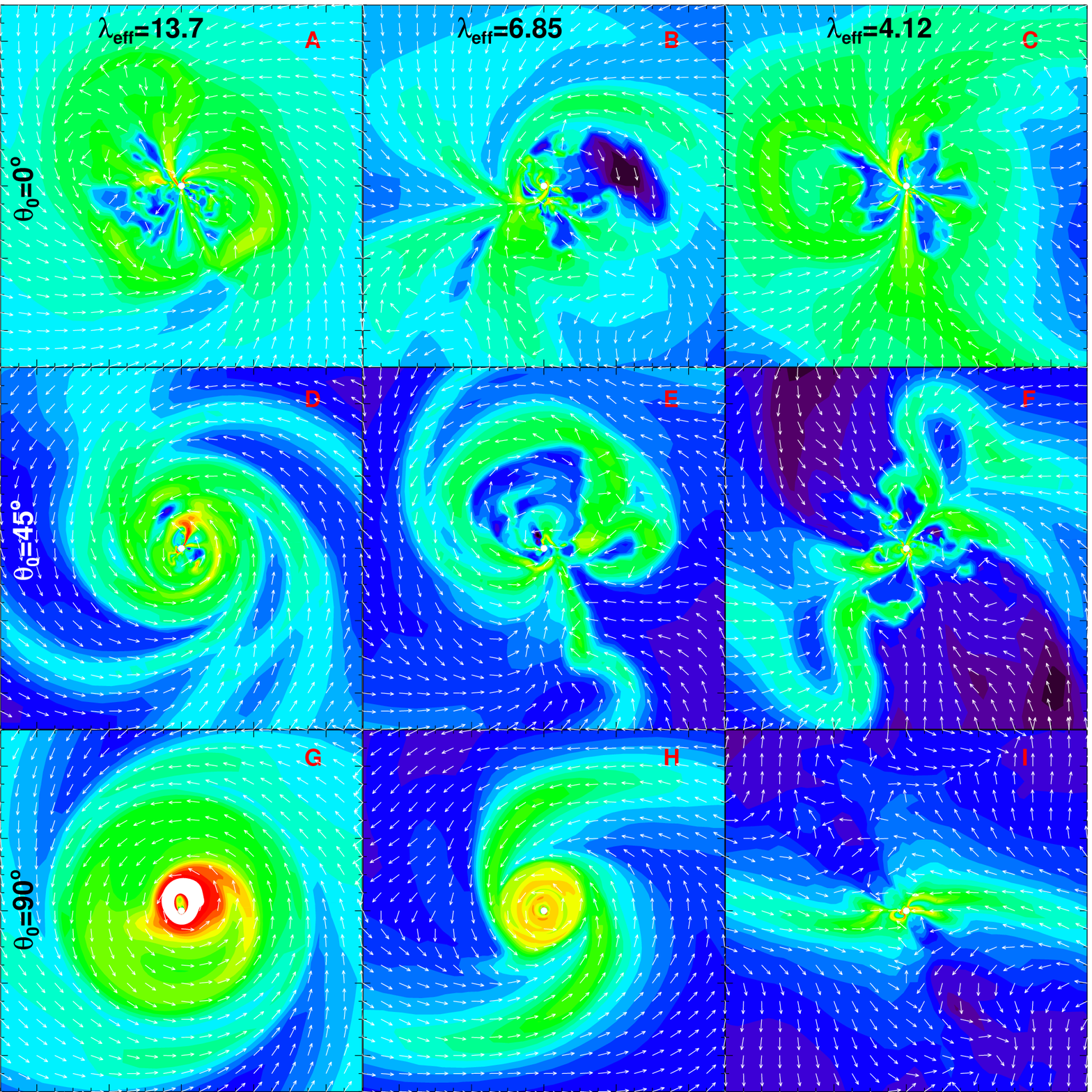}

\caption{Snapshots of Models A--I (see Table~1 for model parameters)
when the central mass reaches $0.2\msun$. Plotted are maps of the
logarithm of density (color range: $-18$ to $-12$) and velocity unit
vectors on the equatorial plane. The box size in each panel is
$10^{16}\cm$. }
\label{all}
\end{figure}

We have seen from the preceding section that, in the weakly magnetized
case of $\lambda=9.72$, the $10^2\AU$-scale inner part of the
protostellar accretion flow is dominated by two very different types
of structures: a weakly magnetized, dense, rotationally
supported disk (RSD) in the orthogonal case
($\theta_0=90\degree$, Model G)
and magnetically dominated, low-density lobes or DEMS in the aligned
case ($\theta_0=0\degree$, Model A), at least at the relatively early
time discussed in \S\ref{misalignment}, when the central mass reaches
$\sim 0.1\msun$. This dichotomy persists to later times
for these two models and for other models as well, as illustrated by
Fig.\ \ref{all}, where we plot Models A--I at a time when the central
mass reaches $0.2\msun$. It is clear that the RSD for Model G
becomes even more prominent at the later
time, although a small magnetically dominated, low-density lobe is
evident close to the center of the disk: it is a trapped DEMS
that is too weak to disrupt the disk. In this case, the
identification of a robust RSD is secure, at even later times (up
to the end of the simulation, when the central mass reaches
$0.38\msun$ or $38\%$ of the initial core mass). In the aligned
case (Model A), the inner accretion flow remains dominated by the
highly dynamic DEMS at late times, with no sign of RSD formation.

For the intermediate tilt angle case of $\theta_0=45\degree$ (Model D),
the inner structure of the protostellar accretion flow is shaped
by the tussle between RSD
and DEMS. Fig.\ \ref{all} shows that, at the plotted time, Model D
has several spiral arms that appear to merge
into a rotating disk. There are, however, at least three
low-density ``holes'' near the center of the disk: they are the
magnetically dominated DEMS. Movies show that the highly variable
DEMS are generally confined close to the center, although they
can occasionally expand to occupy a large fraction of the disk
surface. Overall, the circumstellar structure in Model D is more
disk-like than DEMS-like. We shall call it a ``porous disk,'' to
distinguish it from the more filled-in, more robust disk in the
orthogonal Model G. Even though most of the porous disk has a rotation
speed dominating the infall speed, the infall is highly variable,
and often supersonic. The rotation speed also often deviates greatly
from the Keplerian value. Such an erratic disk is much more dynamic
than the quiescent disks envisioned around relatively mature (e.g., Class
II) YSOs. The intermediate tilt angle case drives a powerful bipolar
outflow, unlike the orthogonal case, but similar to the aligned
case. This is consistent with the rate of angular momentum removal
increasing with decreasing tilt angle (i.e., from $90\degree$ to
$45\degree$; see also \ct{CiardiHennebelle2010}).

As the strength of the initial magnetic field in the core increases,
the DEMS becomes more dominant. This is illustrated in the middle
column of Fig.\ \ref{all}, where the three cases with an intermediate
field strength corresponding to $\lambda=4.86$ (and
$\lambda_{\rm eff}=6.85$) are plotted. In Model H, where the magnetic and
rotation axes are orthogonal, a relatively
small (with radius of $\sim 10^2\AU$) rotationally dominated disk
is clearly present at the time shown. As in the weaker field case
of Model G, it is fed by prominent ``pseudo-spirals'' which are part
of a magnetically-induced curtain in 3D (see the second panel of
Fig.\ \ref{3D}). Compared to Model G, the curtain here is curved to a
lesser degree, which is not surprising because the rotation is slower
due to a more efficient braking and the stronger magnetic field
embedded in the curtain is harder to bend. The disk is also smaller,
less dense,
and more dynamic. It is more affected by DEMS, which occasionally
disrupt the disk, although it always reforms after disruption.
Overall, the circumstellar structure in Model H is more RSD-like
than DEMS-like. As in Model D, we classify it as a ``porous
disk.'' As the tilt angle decreases from $90\degree$ to $45\degree$
(Model E) and further to $0\degree$ (Model B), the rotationally
dominated circumstellar structure largely disappears; it is replaced
by DEMS-dominated structures. Even though there is still a significant
amount of rotation in the accretion flow, a dense coherent disk
is absent. We conclude that for a moderately strong magnetic field
of $\lambda=4.86$ the formation of RSD is suppressed if the tilt angle
is moderate.

\begin{figure}
\epsscale{0.6}
\plotone{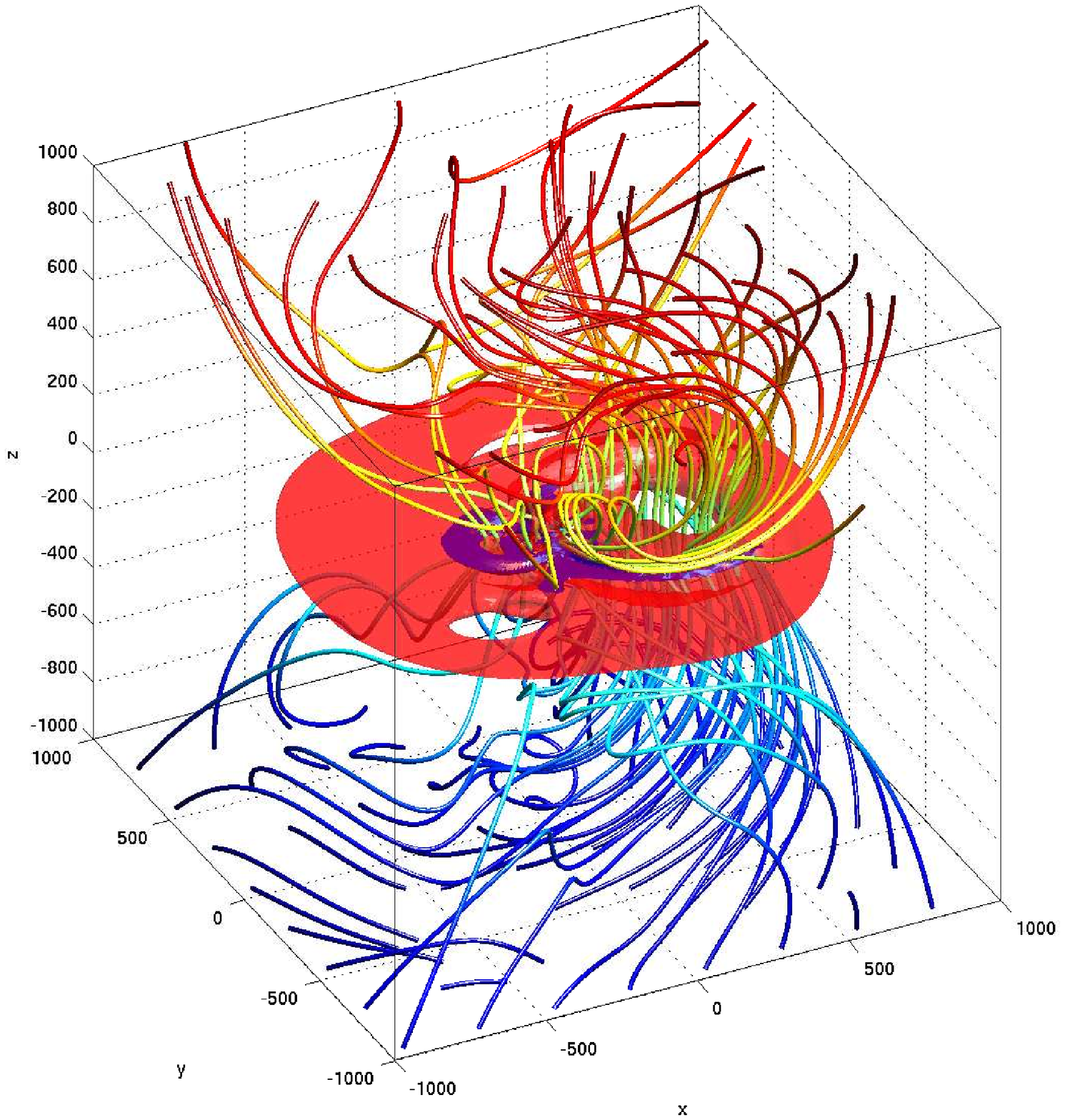}
\plotone{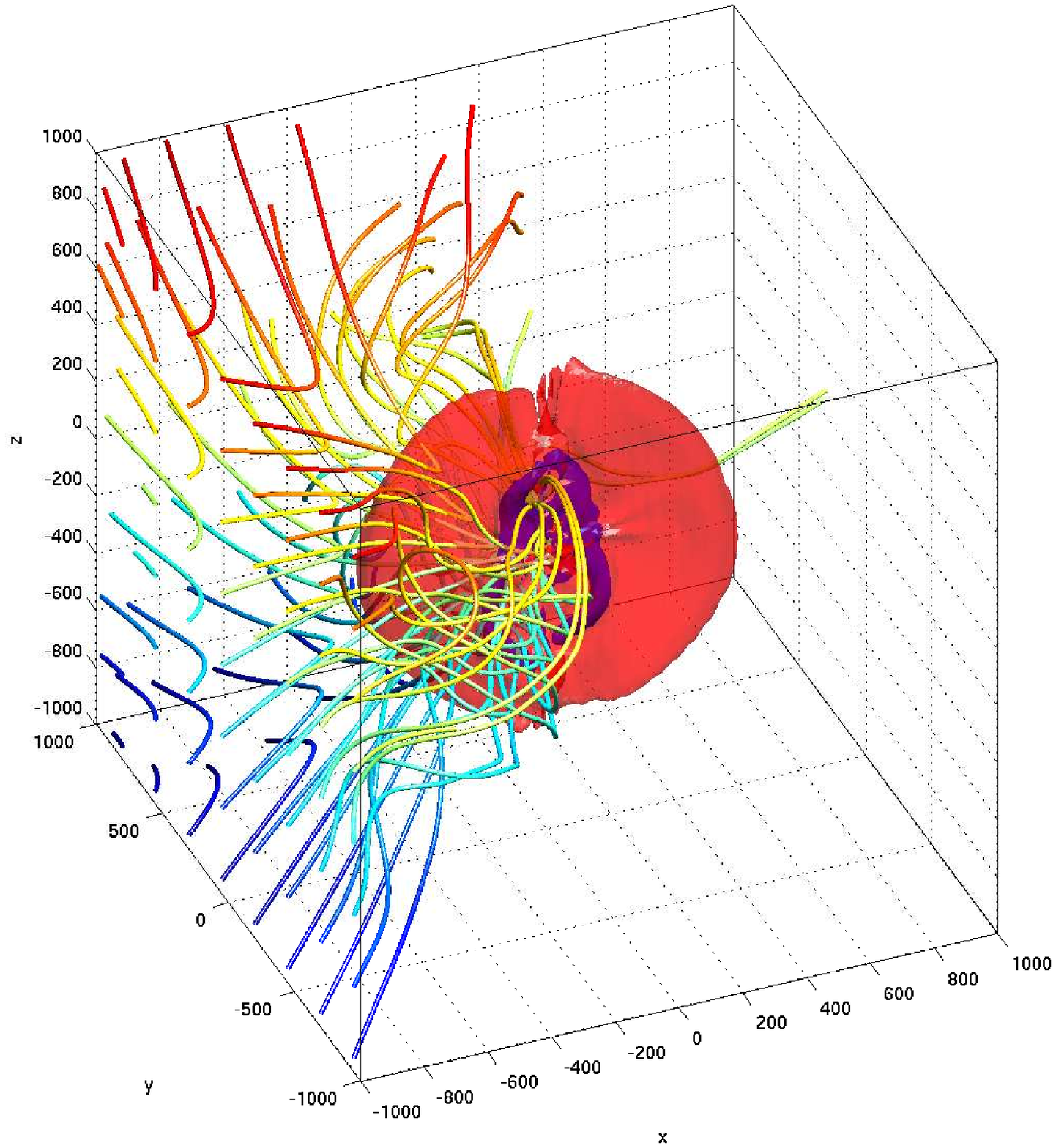}

\caption{Same as in Fig.\ \ref{3D}, but for the stronger field
  ($\lambda=2.92$) cases of
  Model C (aligned) and Model I (orthogonal) at a time when the
  central mass reaches $0.2\msun$.
For clarity, only field lines
  originated from the bottom $x$-$y$ and left $y$-$z$ plane are plotted,
  respectively. Note the difference in orientation of the pseudodisks
  in the two cases. The inner part of both pseudodisks is dominated by
  low-density, strongly magnetized regions. The length is in units of AU.}
\label{3DB1e-5}
\end{figure}

In the cases of the strongest magnetic field corresponding to
$\lambda=2.92$ (and $\lambda_{\rm eff}=4.12$), the formation of
RSD is suppressed regardless of the tilt angle, as can be seen
from the last column of Fig.\ \ref{all}.
For the orthogonal case (Model I), the prominent ``pseudo-spirals''
in the weaker field cases of Model G ($\lambda=9.72$) and H
($\lambda=4.86$) are replaced by two arms that are only slightly
bent. They are part of a well-defined pseudodisk that happens to
lie roughly in the $y=0$ (or $x$-$z$) plane (see the second panel
of Fig.\ \ref{3DB1e-5}). In the absence of any initial rotation,
one would expect the pseudodisk to form perpendicular to the
initial field direction along the $x$-axis, i.e., in the $x=0$
(or $y$-$z$) plane. Over the entire course of core evolution and
collapse, the rotation has rotated the expected plane of
the pseudodisk by nearly $90\degree$. Nevertheless, at the time shown
(when the central mass reaches $0.2\msun$), there is apparently
little rotation left inside $10^3\AU$ to warp the pseudodisk
significantly. Except for the orientation, this pseudodisk looks
remarkably similar to the familiar one in the aligned case
(Model C; see the first panel of Fig.\ \ref{3DB1e-5}). In particular,
there are low-density ``holes'' in the inner part of both pseudodisks
which are threaded by intense magnetic fields and surrounded by
dense filaments: they are the DEMS. In the intermediate tilt angle
case of $45\degree$ (Model F, not shown in the 3D figure), the
pseudodisk is somewhat more warped than the two other cases,
and its inner part is again dominated by DEMS. It is clear that
for a magnetic field of $\lambda$ of a few, the inner circumstellar
structure is dominated by the magnetic field, with rotation playing
a relatively minor role; the RSD remains suppressed despite the
misalignment.

To better estimate the boundary between the cores that produce RSDs
and those that do not, we carried out two additional simulations
with $\theta_0=90\degree$ (Models P and Q in Table~1). We found a
porous disk in Model P ($\lambda=4.03$ and $\lambda_{\rm eff}=5.68$),
as in the weaker field case of Model H, but no disk in Model Q
($\lambda=3.44$ and $\lambda_{\rm eff}=4.85$), as in the stronger
field case of Model I. From this, we infer that the boundary lies
approximately at $\lambda \approx 3.5$ (or $\lambda_{\rm eff}\approx 5$).

\section{Discussion}
\label{discussion}

\subsection{Comparison with \citeauthor{Joos+2012}}

Our most important qualitative result is that the misalignment between
the magnetic field and rotation axis tends to promote the formation of
rotationally supported disks, especially in weakly magnetized
dense cores. This is in agreement with the conclusion previously
reached by \citeauthor{Joos+2012} (\citeyear{Joos+2012}; JHC12 hereafter),
using a different numerical code and
somewhat different problem setup. Their calculations were carried out
using an adaptive mesh refinement (AMR) code in the Cartesian
coordinate system, with the central object treated using a stiffening
of the equation of state, whereas ours were done using a fixed mesh
refinement (FMR) code in the spherical coordinate system, with an
effective sink particle at the origin. Despite the differences, these
two distinct sets of calculations yield qualitatively similar
results. The case for the misalignment promoting disk formation is
therefore strengthened.

Quantitatively, there appears to be a significant discrepancy between
our results and theirs. According to their Fig.\ 14, a Keplerian disk is formed
in the relatively strongly magnetized case of $\lambda=3$ if the
misalignment angle $\theta_0=90\degree$. Formally, this case
corresponds roughly to our Model I ($\lambda=2.92$ and
$\theta_0=90\degree$), for which we can rule out the formation of a
rotationally supported disk with confidence (see Fig.\ \ref{all}).

We believe that the discrepancy comes mostly from the initial density
profile adopted, which affects the degree of magnetization near the
core center for a given global mass-to-flux ratio $\lambda$.
\citeauthor{Joos+2012}
adopted a centrally condensed initial mass distribution
\begin{equation}
\rho (r) = \frac{\rho_0}{1+ (r/r_0)^2}
\label{profile}
\end{equation}
with the characteristic radius $r_0$ set to $1/3$ of the core
radius $R_c$, so that the central-to-edge density contrast is 10
(see also \ct{CiardiHennebelle2010}). It is
easy to show that, for this density profile and a uniform magnetic
field, the mass-to-flux ratio for the flux tube passing through the
origin is
\begin{equation}
\lambda_c = \frac{x_0^2\,\arctan x_0}{2 (x_0 - \arctan x_0)} \lambda
\label{lambda_c}
\end{equation}
where $x_0=R_c/r_0$ and $\lambda$ is the global mass-to-flux ratio for
the core as a whole. For $x_0=3$, we have $\lambda_c=3.21\,\lambda$.
In other words, the central part of their core is substantially less
magnetized relative to mass than the core as a whole, due to the
initial mass condensation. For the $\lambda=3$ case under
consideration, we have $\lambda_c=9.63$,
which makes the material on the central flux tube rather weakly
magnetized (relative to mass). The magnetization of the central region
is important,
because the central part is accreted first and the star formation
efficiency in a core may not be 100\% efficient (e.g., \ct{Alves+2007}).
If we define as in \S\ref{setup} an effective mass-to-flux
ratio for the (cylindrical) magnetic flux surface that encloses
$1/3$ of the core mass, then $\lambda_{\rm eff}=2.39 \lambda$ for
the density distribution adopted by JHC12 (equation [\ref{profile}]).
It is significantly different from the effective mass-to-flux ratio of
$\lambda_{\rm eff}=1.41 \lambda$ for the uniform density that we
adopted.

The above difference in $\lambda_{\rm eff}$ makes our strongest field
case of $\lambda=2.92$ ($\lambda_{\rm eff}=4.12$) more directly
comparable to JHC12's $\lambda=2$ ($\lambda_{\rm eff}=4.78$) case.
There is agreement that, in both cases, the formation of a RSD is
suppressed, even when the tilt angle $\theta_0=90\degree$. These
results suggest that RSD formation is suppressed when the effective
mass-to-flux ratio $\lambda_{\rm eff} \lesssim 5$, independent of
the degree of field-rotation misalignment, consistent with the
conclusion we reached toward the end of \S\ref{strong} based 
on Models P and Q.

Similarly, JHC12's $\lambda=3$ ($\lambda_{\rm eff}=7.17$) models may be
more directly comparable to our $\lambda=4.86$ ($\lambda_{\rm eff}=6.85$)
models. Our calculations show that a (more or less) rotationally
supported disk is formed in the extreme $\theta_0=90\degree$ case
for $\lambda_{\rm eff}=6.85$ (Model H, see Fig.\ \ref{all}) but not in the
intermediate tilt angle $\theta_0=45\degree$ case (Model E). This
is consistent with their Fig.\ 14, where a Keplerian disk is
formed for $\theta_0=90\degree$ for $\lambda_{\rm eff}=7.17$, but not for
$\theta_0=45\degree$. In the latter case, JHC12 found a disk-like
structure with a flat rather than Keplerian rotation profile.
They attributed the flat rotation to additional support from the
magnetic energy, which dominates the kinetic energy at small
radii. We believe that their high magnetic energy comes from
strongly magnetized, low-density lobes (i.e., DEMS), which are
clearly visible in our Model E (see Fig.\ \ref{all}). Although
there is still significant rotation on the $10^2\AU$ scale in
Model E, we find the morphology and kinematics of the circumstellar
structure too disorganized to be called a ``disk.'' We conclude
that for a moderate field strength corresponding to
$\lambda_{\rm eff} \sim 6$--$7$, a rotationally supported disk does not
form except when the magnetic field is tilted nearly orthogonal
to the rotation axis.

Our weak-field case of $\lambda=9.72$ ($\lambda_{\rm eff}=13.7$) can be
compared with JHC12's $\lambda=5$ ($\lambda_{\rm eff}=12.0$) case. In both
cases, a well-defined rotationally supported disk is formed when
$\theta_0=90\degree$ (see our Fig.\ \ref{all} and their Fig.\ 12).
For the intermediate tilt angle $\theta_0=45\degree$, JHC12 obtained
a disk-like structure with a flat rotation curve, with the magnetic
energy dominating the kinetic energy at small radii. This is
broadly consistent with our intermediate tilt angle Model D,
where a highly variable, ``porous'' disk is formed, with the
central part often dominated by strongly magnetized, low-density
lobes. There is also agreement that the disk formation is
suppressed if $\theta_0=0$ even for such a weakly magnetized
case. It therefore appears that, for the weak-field case of
$\lambda_{\rm eff} \gtrsim 10$, a rotationally supported disk can
be induced by a relatively moderate tilt angle of
$\theta_0 \gtrsim 45\degree$.

The result that a misalignment between the magnetic field and
rotation axis helps disk formation by weakening magnetic
braking may be counter-intuitive.
\citet{MouschoviasPaleologou1979}
showed analytically that, for a uniform rotating
cylinder embedded in a uniform static external medium,
magnetic braking is much more efficient in the orthogonal
case than in the aligned case. This analytic result may not
be directly applicable to a collapsing core, however. As
emphasized by JHC12, the collapse drags the initially uniform,
rotation-aligned magnetic field into a configuration that
fans out radially. JHC12 estimated analytically that the
collapse-induced field fan-out could in principle make the
magnetic braking in the aligned case more efficient than in
the orthogonal case. The analytical estimate did not, however,
take into account of the angular momentum removal by outflow,
which, as we have shown in \S\ref{torque}, is a key
difference between the weak-field ($\lambda=9.72$, $\lambda_{\rm
  eff}=13.7$) aligned
case (Model A) where disk formation is suppressed and its
orthogonal counterpart (Model G) that does produce a rotationally
supported disk (see Figs.\ \ref{contrast} and \ref{Torque},
and also \ct{CiardiHennebelle2010}).

The generation of a powerful outflow in the weak-field aligned case
(Model A) is facilitated by the orientation of its pseudodisk,
which is perpendicular to the rotation axis (see Fig.\ \ref{3D}).
This configuration is conducive to both the pseudodisk winding up
the field lines and the wound-up field escaping above and below the
pseudodisk, which drives a bipolar outflow. When the magnetic field
is tilted by $90\degree$ away from the rotation axis, the pseudodisk
is warped by rotation into a snail-shaped curtain that is
unfavorable to outflow driving (see Fig.\ \ref{3D}). The outflow
makes it more difficult to form a rotationally supported disk in
the aligned case. Disk formation is further hindered by magnetically
dominated, low-density lobes (DEMS), which affect the inner part
of the accretion flow of the aligned case more than that of the
perpendicular case, at least when the field is relatively weak
(see Fig.\ \ref{contrast} and \ref{all}). For more strongly
magnetized cases, the DEMS becomes more dynamically important
close to the central object, independent of the tilt angle
$\theta_0$ (see Fig.\ \ref{3DB1e-5}). DEMS-like structures were
also seen in some runs of JHC12 (e.g., the case of $\lambda=2$
and $\theta_0=0\degree$; see their Fig.\ 19) but were not commented
upon. As stressed previously by \citet{Zhao+2011} and
\citet{Krasnopolsky+2012} and confirmed by our calculations,
the DEMS presents a formidable obstacle to the formation and
survival of a rotationally supported disk.

\subsection{Misalignment not Enough for Disk Formation in General}
\label{disk}

While there is agreement between JHC12 and our calculations that
misalignment between the magnetic field and rotation axis is
beneficial to disk formation, it is unlikely that the misalignment
alone can enable disk formation around the majority of young
stellar objects. The reason is the following.
\citet{TrolandCrutcher2008}
obtained a mean dimensionless mass-to-flux ratio of
${\bar \lambda_{\rm eff}}\approx 2$ through OH Zeeman
observations for a sample of
dense cores in nearby dark
clouds\footnote{We assume that the mass-to-flux ratio
measured by Troland \& Crutcher on the core scale is the
same as the effective mass-to-flux ratio near the core
center because, for significantly magnetically supercritical
cores, ambipolar diffusion is generally ineffective in
reducing the value of $\lambda$ near the center relative
to that in the envelope.}.
For such a small $\lambda_{\rm eff}$, disk formation is completely
suppressed, even for the case of maximum misalignment of
$\theta_0=90\degree$.
\citet{Crutcher+2010} argued, however, that there is a
flat distribution of the total field strength in dense cores,
from $B_{\rm tot}\approx 0$ to some maximum value
$B_{\rm max}\approx 30\muG$;
the latter corresponds to $\lambda_{\rm eff} \approx 1$, so that
the mean ${\bar \lambda_{\rm eff}}$ stays around 2. If this is the
case, some cores could be much more weakly magnetized than others,
and disks could form preferentially in these cores.
However, to form a rotationally supported disk,
the core material must have
(1) an effective mass-to-flux ratio $\lambda_{\rm eff}$ greater
than about 5 {\it and} (2) a rather large tilt angle
(see discussion in the preceding section). If one assumes
that the core-to-core variation of $\lambda_{\rm eff}$ comes mostly
from the field strength
rather than the column density (as done in \ct{Crutcher+2010}),
then the probability of a core having $\lambda_{\rm eff} \gtrsim 5$
(or $B_{\rm tot}\lesssim 6\muG$) is $\sim 1/5$. Since a large tilt
angle of $\theta_0\sim 90\degree$ is required to form a RSD for
$\lambda_{\rm eff} \sim 6$--$7$, the chance of disk formation is reduced
from $\sim 1/5$ by at least a factor of 2 (assuming a random
orientation of the magnetic field relative to the rotation axis),
to $\sim 10\%$ or less.

The above estimate is necessarily rough, and can easily be off by
a factor of two in either direction. It is, however, highly unlikely
for the majority of the cores to simultaneously satisfy the
conditions on both $\lambda_{\rm eff}$ and $\theta_0$ for disk formation.
The condition $\lambda_{\rm eff} \gtrsim 5$ is especially difficult to
satisfy because, as noted earlier, it implies that the dense cores
probed by OH observations must have a total field strength $B_{\rm tot}
\lesssim 6\muG$, less than or comparable to the well-defined median
field strength inferred by \citet{HeilesTroland2005} for the
much more diffuse, cold neutral {\it atomic} medium (CNM). It
is hard to imagine a reasonable scenario in which the majority of
dense cores have magnetic fields weaker than the CNM.

We note that \citet{Krumholz+2013} independently estimated a
range of $\sim 10$--$50\%$ for the fraction of dense cores
that would produce a Keplerian disk based on Fig.\ 14 of JHC12.
Their lower limit of $\sim 10\%$ is in agreement with our
estimate; in both cases, the fraction is dominated by weakly
magnetized cores that have mass-to-flux ratios greater than $5$
and moderately large tilt angles. Their upper limit of
$\sim 50\%$ is much higher than our estimate, mainly because
it includes rather strongly magnetized cores with mass-to-flux
ratios as small as $2$. Since our calculations show that
such strongly magnetized cores do not produce rotationally
supported disks even for large tilt angles, we believe that
this upper limit may be overly generous.

Whether dense cores have large tilt angle $\theta_0$ between
the magnetic field and rotation axis that are conducive to disk
formation is unclear. \citet{Hull+2012} measured the field
orientation on the $10^3\AU$-scale for a sample of 16 sources
using millimeter interferometer CARMA. They found that the field
orientation is not tightly correlated with the outflow axis;
indeed, the angle between the two is consistent with being
random. If the outflow axis is aligned with the core rotation
axis and if the field orientation is the same on the core
scale as on the smaller, $10^3\AU$-scale, then $\theta_0$
would be randomly distributed between $0$ and $90\degree$, with
half of the sources having $\theta_0 \geq 60\degree$. However,
the outflow axis may not be representative of the core rotation
axis. This is because the (fast) outflow is thought to be driven
magnetocentrifugally from the inner part of the circumstellar
disk (on the AU-scale or less;
\ct{Shu+2000}; \ct{KoniglPudritz2000}).
A parcel of core material would have lost most
of its angular momentum on the way to the outflow launching
location; the torque (most likely magnetic or gravitational)
that removes the angular momentum may also change the direction of the
rotation axis. Similarly, the field orientation on the $10^3\AU$-scale
may not be representative of the initial field orientation on
the larger core scale. The magnetic field on the $10^3\AU$
scale is more prone to distortion by collapse and rotation
than that on the core scale. Indeed, \citet{Chapman+2013}
found that the field orientation on the core scale measured
using single-disk telescope CSO is within $20\degree$ of the
outflow axis for 3 of the 4 sources in their sample; the larger
angle measured in the remain source may be due projection
effects because its outflow axis is close to the line-of-sight.
If the result of \citeauthor{Chapman+2013} is valid in general
and if the
outflow axis reflects the core rotation axis, then dense cores
with large tilt angle $\theta_0$ would be rare. In this case,
disk formation would be rare according to the calculations
presented in this paper and in JHC12, even in the unlikely event
that the majority of dense cores are as weakly magnetized as
$\lambda_{\rm eff} \gtrsim 5$.

Rotationally supported disks are observed, however, routinely
around evolved Class II YSOs (see \ct{WilliamsCieza2011}
for a recent review), and increasingly around younger Class
I (e.g., \ct{Jorgensen+2009}; \ct{Lee2011}; Takakuwa et al. 2012) and even Class 0
(\ct{Tobin+2012}) sources. When and how such disks form remain
unclear. Our calculations indicated that the formation of large
observable ($10^2\AU$-scale) rotationally supported disks\footnote{We
cannot rule out the existence of small, AU-scale disks, because the
size of our effective ``sink particle'' is $6.7\AU$. Small
disks may be needed to drive fast outflows during the Class 0
phase, and may form through non-ideal MHD effects
(\ct{Machida+2010}; \ct{DappBasu2010}; \ct{Dapp+2012}).}
is difficult even in the presence of a large tilt angle
$\theta_0$ during the early protostellar accretion (Class 0)
phase\footnote{We are unable to follow the collapse until most
of the core material is accreted into the central object for
numerical reasons. Such a complete accretion may not be
realistic, however, if the star formation efficiency in a
core is typically as low as $1/3$ (e.g., \ct{Alves+2007}).}.
This may not contradict the available
observations yet (\ct{Maury+2010}), since only one Keplerian
disk is found around a (late) Class 0 source so far (L1527;
\ct{Tobin+2012}); it could result from a rare combination of
an unusually weak magnetic field and a large tilt angle $\theta_0$.
If it turns out, through future observations using ALMA and
JVLA, that large-scale Keplerian disks are prevalent around
Class 0 sources, then some crucial ingredients must be missing
from the current calculations. Possible candidates include
non-ideal MHD effects and turbulence. The existing calculations
indicate that realistic levels of the classical non-ideal MHD
effects do not weaken the magnetic braking enough to enable
large-scale disk formation (\ct{MellonLi2009}; \ct{Li+2011};
\ct{Krasnopolsky+2012}; see also \ct{KrasnopolskyKonigl2002} and
\ct{BraidingWardle2012a,BraidingWardle2012b}),
although misalignment has yet to be
considered in such calculations. Supersonic turbulence was found
to be conducive to disk formation (\ct{Santos-Lima+2012,Santos-Lima+2013};
\ct{Seifried+2012}; \ct{Myers+2012}; \ct{Joos+2013}), although dense cores
of low-mass star formation typically have subsonic non-thermal
line-width and it is unclear whether subsonic turbulence can
enable disk formation in dense cores magnetized to a realistic
level. If, on the other hand, it turns out that large-scale
Keplerian disks are rare among Class 0 sources, then the question
of disk growth becomes paramount: how do the mostly undetectable
Class 0 disks become detectable in the Class I and II phase?
If the magnetic braking plays a role in keeping the early disk
undetectable, then its weakening at later times may promote
rapid disk growth. One possibility for the late weakening of
magnetic braking is the depletion of the protostellar envelope,
either by outflow stripping (\ct{MellonLi2008}) or accretion
(\ct{Machida+2011}).
It deserves to be better quantified.

\subsection{Summary}
\label{summary}

We carried out a set of MHD simulations of star formation in dense
cores magnetized to different degrees and with different tilt angles
between the magnetic field and the rotation axis. We confirmed the
qualitative result of \citet{Joos+2012} that misalignment between
the magnetic field and rotation axis tends to weaken magnetic
braking and is thus conducive to disk formation. Quantitatively,
we found however that the misalignment enables the formation of a
rotationally supported disk only in dense cores where the star-forming
material is rather weakly magnetized, with a dimensionless mass-to-flux
ratio $\gtrsim 5$; large misalignment in such cores allows the
rotation to wrap
the equatorial pseudodisk in the aligned case into a curved curtain
that hinders outflow driving and angular momentum removal, making
disk formation easier. In more strongly magnetized cores, disk
formation is suppressed independent of the misalignment angle, because
the inner part of the protostellar accretion flow is dominated by
strongly magnetized, low-density regions. If dense cores are as
strongly magnetized as inferred by \citeauthor{TrolandCrutcher2008}
(\citeyear{TrolandCrutcher2008}; with a mean mass-to-flux ratio
$\sim 2$), it would be difficult for the misalignment alone to enable
disk formation in the majority of them. We conclude that how
protostellar disks form remains an open question.

We thank Mark Krumholz for useful discussion. This work is supported in part by NASA grant NNX10AH30G and the
Theoretical Institute for Advanced Research in Astrophysics (TIARA) in
Taiwan through the CHARMS project.

\end{document}